\documentclass[%
 reprint,
 amsmath,amssymb,
 aps,
]{revtex4-2}

\usepackage{graphicx}
\usepackage{dcolumn}
\usepackage{bm}

\begin{document}

\preprint{APS/123-QED}

\title{When low-loss paths make a binary neuron trainable: detecting algorithmic transitions with the connected ensemble}

\author{Damien Barbier}%
 \email{damien.barbier@unibocconi.it}
\affiliation{%
 Department of Computing Sciences, BIDSA, Bocconi University, Milan, MI 20100, Italy
}%

\begin{abstract}
We study the \emph{connected} ensemble, a statistical-mechanics framework that characterizes the formation of low-loss paths in rugged landscapes. First introduced in \cite{barbier2025findingrightpathstatistical}, this ensemble allows one to identify when a network can be trained on a simple task and which minima should be targeted during training. We apply this new framework to the symmetric binary perceptron model (SBP), and study how its typical {connected} minima behave.
We show that {connected} minima exist only above a critical threshold $\kappa_{\rm connected}$, or equivalently below a critical constraint density $\alpha_{\rm connected}$. This defines a parameter range in which training the network is easy, as local algorithms can efficiently access this {connected} manifold. We also highlight that these minima become increasingly robust and closer to one another as the task on which the network is trained becomes more difficult.
\end{abstract}

\maketitle


\section{Introduction}

Once confined to a handful of theoretical models,  the characterization of systems with a rugged loss landscape has become a ubiquitous challenge in recent years. It now spans a wide range of disciplines, encompassing diverse phenomena: from our understanding of  neural networks and their learning capabilities to the  phylogenetics of biological systems. With different perspectives, each field has unveiled important breakthroughs: the glassy transition and aging dynamics in disordered physics \cite{mezard1987spin,parisi1979infinite,PhysRevLett.71.173}, worst-case complexity and locally stable algorithms in computer science \cite{gamarnik2021survey,Achlioptas2011,gamarnik2014limits} or even the inference of phylogenetic trees in biology \cite{Rannala1996,Huelsenbeck2001}. The main drawback of these advances is usually their reliance on strong statistical-mechanics postulates. They typically describe systems at equilibrium, assuming ergodicity and the principle of indifference (or entropy maximization). However, over the years,  dynamical systems that actually fall within this statistical-mechanics framework have become the exception rather than the norm.
Neural networks (NN), virus dynamics -to name a few- are the new setups of interest, and they do not reach the standard Gibbs-Boltzmann equilibrium implied with these postulates. Understanding in which regions such setups can navigate remains in general an open problem.

Interestingly, although often unnoticed, a common picture seems to emerge from three very different on-going research directions: machine learning, biophysics, and computer science. Whether it is the training of a neural network, the evolution process of a virus, or stable algorithms for a constraint satisfaction problem, dynamics/algorithms always appear to converge towards unexpectedly flat regions in these problems' loss landscapes \cite{Barbier2025,Mauri2023,baldassi2021unveiling,han2025flatnessnecessaryneuralcollapse}. In other words, these landscapes contain strong dynamical attractors formed by aggregations of minima.

To understand the formation of such attractors, toy models with rugged landscapes have been designed and thoroughly studied; lattice protein model in biophysics \cite{Lau1989,Lat_prot,Mauri2023}, committee machine in machine learning \cite{Tresp2000BCM}, or quantum spin glass in physics \cite{PhysRevResearch.3.043015,Baldassi2018} to name a few. Regardless of the model, the overall setup remains the same: a system with a large number of degrees of freedom has to navigate in a complex landscape, its dynamics only allowing for incremental reconfigurations. 
In the plethora of available toy models, a case of particular interest is the perceptron model. Often presented as a one-layer NN, it consists of an optimization problem in which a $N$-dimensional vector ${\bf x}$ has to correctly classify an extensive number of random data points $\{\xi^\mu\}_{\mu \in [\![1,M]\!]}$ (for which extensive means $M\propto N$) given an activation function. Since its first appearance in statistical physics in the 1980s \cite{gardner1988optimal}, many variants of this model have been scrutinized with different phase spaces for ${\bf x}$, data point distributions or activation functions \cite{Annesi2023,krauth89storage,baldassi2020clustering,aubin2019storage}. Its main interest lies in its relative simplicity compared to the models mentioned above; while still exhibiting the same phenomenology of flat dynamical attractors. 

For this model, the discrepancy between standard landscape characterization and dynamics can be summarized as follows. On the one hand, using statistical mechanics tools to analyze the landscape, it has been shown that typical solutions to the perceptron task (a.k.a.\ equilibrium solutions) cannot be reached in polynomial time by local algorithms.
This is because these typical solutions are in fact surrounded by a sea of poorly performing configurations.
In the literature, these are often referred to as \emph{isolated} solutions \cite{krauth89storage} or as exhibiting the overlap-gap property \cite{gamarnik2021survey}.
To reach such solutions, an algorithm would have to jump over wide regions of poorly performing configurations. Realistic learning routines, which can only perform local updates, are incapable of making such long jumps efficiently. 
On the other hand, modern algorithms are now capable of returning viable solutions to the perceptron task. These new local algorithms appear to be insensitive to the abundance of \emph{isolated} solutions in the loss landscape. In fact, numerous works \cite{baldassi2016local,baldassi2015subdominant,baldassi2020clustering} have pointed out that such algorithms actually target atypically flat regions.
In other words, they target regions where many solutions can be found close to each other.
This leaves one major problem: the statistical mechanics tools at our disposal are incapable of describing such atypical manifolds.

In this direction, a recent breakthrough has been the introduction of the \emph{connected} ensemble \cite{Barbier2025,barbier2025findingrightpathstatistical}. It is an approach based on a simple idea: studying how paths of minima are distributed in the perceptron loss landscape. If typical solutions are \emph{isolated} (and therefore inaccessible), a solution path $\{{\bf x}^k\}_{k\in [\![0, k_f]\!]}$ should, with high probability, lie in flat (and accessible) regions.
This is because each configuration ${\bf x}^k$ lies close to its direct neighbors ${\bf x}^{k\pm 1}$.
Therefore, this path structure contains at least one flat direction, namely along the path itself.
Local algorithms can then access these manifolds because the solutions they contain are not separated from each other by wide regions composed of poorly performing configurations.
First studied in \cite{barbier2025findingrightpathstatistical}, this new statistical ensemble unveils the structure of the flat manifolds present in the perceptron loss landscape. More practically, it also led to the design of modified Monte Carlo algorithms that access these flat regions. The main limitation of the description proposed in \cite{barbier2025findingrightpathstatistical} is its reliance on the \emph{no-memory Ansatz}.
In a few words, this hypothesis allows one to carry out analytical computations by fixing the geometry of the solution paths. Although these \emph{no-memory} paths exist and are accessible, they do not correspond to the most numerous paths in the perceptron loss landscape.

In this paper, we will focus on characterizing flat manifolds beyond the \emph{no-memory Ansatz}. To maintain continuity with \cite{barbier2025findingrightpathstatistical}, we study the same model, namely the symmetric binary perceptron (SBP). This will allow us to address the differences between the more general flat manifolds and the \emph{no-memory} ones. The remainder of this paper is organized as follows. In Sec.~\ref{sec: defs}, we introduce the perceptron model and define -in a generic form- the statistical ensemble for connected minima. In Sec.~\ref{sec: analysis}, we analyze the saddle point of the \emph{connected} free energy. In particular, we will address the geometry of the paths, their robustness, and the algorithmic thresholds for training the SBP.

\section{Definitions and connected free energy}
\label{sec: defs}
\subsection{The model}
The symmetric binary perceptron (SBP) is a setup that was first introduced in \cite{aubin2019storage}. It corresponds to a classification task performed by a binary neuron. In more detail, its training consists of finding a $N$-dimensional binary vector ${\bf x} \in \{-1,+1\}^N$ that is capable of satisfying an extensive number of random inequalities. These random inequalities are constructed as follows. Given an ensemble of $M$ i.i.d. random patterns $\{\xi^\mu\}_{\mu\in[\![1,M]\!]}$, with $\xi^\mu\in {\rm I\!R}^N$, we want to find ${\bf x}$ on the hypercube (${\bf x}\in \Sigma^N$) such that
\begin{align}
\label{eq: constraints}
    \vert\xi^\mu\cdot{\bf x}\vert\leq\kappa\sqrt{N} ~~~~~ \mbox{for all}~ 1 \le \mu \le M\, .
\end{align}
For simplicity, we choose normal distributed patterns, i.e. $\xi^\mu\sim\mathcal{N}(0,{\bf I})$. 
This problem is often rephrased as finding a minimum $\bf x$ in the SBP loss landscape:
\begin{align}
    {\bf x}\in{\rm argmin}_{S^N} \mathcal{L} 
\end{align}
with
\begin{align}
    \mathcal{L}({\bf x})=\sum_{\mu=1}^M \Theta\left(\left\vert \frac{ \xi^\mu \cdot {\bf x}}{\sqrt{N}}\right\vert-\kappa\right) \, .
\end{align}
The loss function $\mathcal{L}(\cdot)$ is always positive and becomes null only when $\bf x$ verifies all the constraints. We denote here the Heaviside function by $\Theta(\cdot)$.

In the $N\rightarrow+\infty$ limit, this model is controlled by two positive parameters: the constraint density $\alpha=M/N$ and the threshold $\kappa$. Straightforwardly, finding a solution to this task becomes increasingly more difficult as $\alpha$ increases and vice versa as $\kappa$ decreases. More precisely, it was proven in ~\cite{aubin2019storage} that this problem admits solutions with high probability if and only if the margin threshold $\kappa$ verifies
\begin{align}
    \kappa>\kappa^\alpha_{\rm SAT}
\end{align}
with
\begin{align}
\label{eq:SAT}
    \log(2)+\alpha\log\left(\int \mathcal{D}u\,\Theta(\kappa^\alpha_{\rm SAT}-\vert u\vert)\right)=0\,.
\end{align}
Throughout this paper, we will use the notation $\mathcal{D}.$ to represent an integration with a scalar normal-distributed variable. As it was first described in the seminal paper of M\'ezard and Krauth \cite{krauth89storage}, the manifold of solutions is dominated in number by \emph{isolated} configurations. In other words, typical minima ${\bf x}_{\rm typ}$ of the loss function $\mathcal{L}(\cdot)$ are surrounded by a sea of high-loss configurations $\bf x$ (i.e. $\mathcal{L}({\bf x})\gg 1$). This is an indirect consequence of their geometrical structure; sometimes called {\it frozen replica symmetry breaking} \cite{martin2004frozen,zdeborova2008locked,huang2013entropy,huang2014origin,Semerjian}. Such \emph{isolated} minima are known to be inaccessible to local algorithms in polynomial time due to the presence of these surrounding high-loss barriers. It is known as the overlap-gap property (OGP) \cite{gamarnik2021survey}. Further investigations of the SBP have shown that, although certain atypical and robust solutions are not \emph{isolated} (that is, they are surrounded by other low-loss configurations) \cite{barbier2023atypical}, simple dynamics such as standard Monte Carlo algorithms fail to navigate these regions on short time scales \cite{Barbier2025}.

Despite these difficulties in navigating the SBP loss landscape, sophisticated algorithms are now capable of returning atypical solutions within a finite range of parameters $\alpha$ and $\kappa$ \cite{bansal2020line,baldassi2016local,braunstein2006learning}. Over the years, several works have attributed this success to the presence of so-called atypical \emph{dense} regions of solutions \cite{baldassi2016unreasonable,baldassi2016local,baldassi2015subdominant,baldassi2020clustering,baldassi2021unveiling}. These regions would correspond to a subset of minima lying at a finite Hamming distance from each other, i.e. minima sharing a finite fraction of their binary entries. In fact, the study in \cite{barbier2025findingrightpathstatistical} has shown that the algorithmic transition for the SBP is due to the presence of solution paths in the loss landscape. These paths form a navigable manifold that can be targeted by local algorithms. The algorithmic transition is then dictated by whether these navigable regions are surrounded by entropic or loss barriers. When such barriers are present, local algorithms with random initialization cannot enter these regions: training the SBP is hard. When there are no barriers, local algorithms can find these flat manifolds and the network can be trained. In \cite{barbier2025findingrightpathstatistical}, it was shown that these barriers appear only for $\kappa < \kappa^{\rm no\text{-}mem}_{\rm loc.\,stab.}$ (with $\alpha$ fixed and paths following the \emph{no-memory Ansatz}). The question that remains is whether paths with a different geometry can generate more navigable regions.

Before moving to the definition of the \emph{connected} free energy, let us introduce three essential quantities for the study of the SBP. The first one is the margins (given a configuration $\bf x$)
\begin{align}
    w^\mu=\frac{\xi^\mu\cdot{\bf x}}{\sqrt{N}}\quad \mbox{for}\quad \mu\in[\![1,M]\!]
\end{align}
and more particularly their statistical distribution $P(w)$. Emphasized in several works \cite{Barbier2025,barbier2023atypical}, this quantity quantifies how well {\bf x} classify each data point. The second quantity are overlaps. Namely, if we take two configurations ${\bf x}^a$ and ${\bf x}^b$ their overlap is defined as
\begin{align}
    m=\frac{{\bf x}^a \cdot {\bf x}^b}{N}\, .
\end{align}
In high dimensions, it turns out (as we will see) that a configuration path $\{{\bf x}^{k}\}_{k\in[\![0,k_f]\!]}$ can be fully described by the overlap matrix \cite{PhysRevLett.71.173,Barbier2025,Franz_2013}
\begin{align}
    {\bf Q}_{k,k'}=\frac{{\bf x}^{k}\cdot {\bf x}^{k'}}{N}\, .
\end{align}
Finally, the last quantity we have to introduce is the partition function $\mathcal{Z}$  for the solution of the perceptron problem (with its associated free energy $\phi$):
\begin{align}
\label{eq: partition func def}
    \mathcal{Z}=\sum_{{\bf x}\in \Sigma^N}e^{\mathcal{L}_{\rm SBP}({\bf x})}\,~~\text{with}~~ \phi=\log{\mathcal{Z}}
\end{align}
and
\begin{align}
\label{eq: SBP loss}
   e^{\mathcal{L}_{\rm SBP}({\bf x})}=\prod_{\mu =1}^M\Theta\left(\kappa-\left\vert\frac{\xi^\mu\cdot {\bf x}}{\sqrt{N}}\right\vert\right) \, .
\end{align}
Thus, if a configuration $\bf x$ does not verify all the constraints from Eq.~(\ref{eq: constraints}), it will not contribute in the SBP partition function ($e^{\mathcal{L}_{\rm SBP}({\bf x})}=0$) . Conversely, any configurations solving the problem are given the same weight ($e^{\mathcal{L}_{\rm SBP}({\bf x})}=1$). As mentioned above, this partition function is in fact dominated by \emph{isolated} solutions. A standard method to show this is to evaluate with the replica method its disordered-averaged free energy \cite{gardner1988optimal,krauth89storage,aubin2019storage}
\begin{align}
    \phi={\rm I\!E}_{\xi}[\log(\mathcal{Z})]\underset{n\rightarrow 0}{=}\frac{{\rm I\!E}_{\xi}[\mathcal{Z}^n]-1}{n}
\end{align}
where ${\rm I\!E}_{\xi}$ indicates the average over the patterns distribution. We will use the same technique later to compute partition function of the \emph{connected} ensemble.

\subsection{The connected free energy}
As introduced in \cite{barbier2025findingrightpathstatistical}, the \emph{connected} ensemble counts the solutions ${\bf x}^0$ that belong to a connected manifold. In other words, it counts the configurations ${\bf x}^0$ that solve the SBP task under the additional requirement that other solutions 
${\bf x}^1$ exist in their immediate vicinity (i.e., with ${\bf x}^0\cdot {\bf x}^1/N=m$ and $m\approx 1$). These minima ${\bf x}^1$, which are themselves connected, have other neighboring solutions ${\bf x}^2$, again satisfying $\mathbf{x}^1 \cdot \mathbf{x}^2 / N = m$.
This construction continues iteratively so that any solution ${\bf x}^k$ (with $k\in[\![1,k_f-1]\!]$) has a nearby solution ${\bf x}^{k+1}$ with ${\bf x}^k\cdot {\bf x}^{k+1}/N=m$.  The partition function associated with such a construction is
\begin{align}
\label{eq: connected measure}
    \mathcal{Z}=\sum_{{\bf x}^0\in \Sigma^N}e^{-\mathcal{L}_{\rm SBP}\left({\bf x}^0\right)+y_1\phi_1\left({\bf x}^0,m\right)}
\end{align}
with the biases ($k\in[\![1,k_f-1]\!]$)
\begin{align}
    \hspace{-0.115cm}\phi_k\left({\bf x}^{k-1},m\right)\!=\!
    \log\!\left[\!\sum_{{\bf x}^k\in \Sigma^{N,m}_{{\bf x}^{k-1}}}\hspace{-0.25cm}e^{-\mathcal{L}_{\rm SBP}\left({\bf x}^k\right)+y_{k+1}\phi_{k+1}\left({\bf x}^{k},m\right)}\!\right]
\end{align}
and
\begin{align}
    \phi_{k_f}\left({\bf x}^{k_f-1},m\right)=
    \log\left[\sum_{{\bf x}^{k_f}\in \Sigma^{N,m}_{{\bf x}^{k_f-1}}}e^{-\mathcal{L}_{\rm SBP}\left({\bf x}^{k_f}\right)}\right].
\end{align}
For compact notation, we use $\Sigma^{N,m}_{{\bf x}^*}$ to represent the space of binary configurations $\bf x$ that verify ${\bf x}\cdot {\bf x}^*/N=m$. In the following, we will focus exclusively on the case $y_k=1$ (for $k\in[\![0,k_f]\!]$). This will allow us to narrow down our analysis to one type of connected manifold, where each configuration ${\bf x}^k$ interacts exactly with one left-neighbor ${\bf x}^{k-1}$ and one right-neighbor ${\bf x}^{k+1}$. In this special case, the partition function reads
\begin{align}
    \mathcal{Z}=\sum_{{\bf x}^0\in \Sigma^N}e^{-\mathcal{L}_{\rm SBP}({\bf x}^0)}\prod_{k=1}^{k_f}\sum_{{\bf x}^k\in \Sigma^{N,m}_{{\bf x}^{k-1}}}e^{-\mathcal{L}_{\rm SBP}({\bf x}^k)}\, .
\end{align}

As a reminder, the SBP loss involves random patterns -see Eq.~(\ref{eq: SBP loss})-. This means that we will have to add an average over the patterns' distribution. In this case, the standard quantity to evaluate is the associated $quenched$ free energy
\begin{align}
    \phi={\rm I\!E}_\xi\left[\log(\mathcal{Z})\right]\, .
\end{align}
In our case, the loss function is actually symmetric under the transformation ${\bf x}\rightarrow-\bf x$. Therefore, and as already used in \cite{barbier2025findingrightpathstatistical}, we can assume that the free energy can be evaluated at the \emph{annealed} level:
\begin{align}
     \phi\approx\log[{\rm I\!E}_\xi\left(\mathcal{Z}\right)]\,.
\end{align}
This approximation is exact if two typical solutions ${\bf x}^{0,a}$ and  ${{\bf x}^{0,b}}$ -sampled from the measure defined in Eq.~(\ref{eq: connected measure})- do not overlap with each other, i.e. ${\bf x}^{0,a}\cdot{{\bf x}^{0,b}}/N=0$. As we will see later, this approximation is correct. 

Leaving the detail of the computation in App.~\ref{app: General comp}, the free energy reads
\begin{align}
\label{eq: free energy}
    \phi&\!=\!\!\underset{
    \tiny\begin{array}{c}
         \hat{\bf Q}_{k,k'}  \\
         {\bf Q}_{k,k'}\\
           ({\bf Q}_{k,k\pm1}\!=\!m)
    \end{array}}{\rm opt}\!\!\!\!\!\left\{\!N\log\!\left(\mathcal{Z}_S\right)+M\log\!\left(\frac{\mathcal{Z}^\kappa_{E}}{\mathcal{Z}^{+\infty}_{E}}\right)\!\right\} \, ,\\
    \label{eq: Z entropy}
    \mathcal{Z}_S&\!=\!\prod_{k=0}^{k_f}\sum_{{x}^{{k}}=\pm 1}e^{\underset{k,{k'}}{\sum}\!\hat{\bf Q}_{k,k'}\left({x}^{{k}}{ x}^{{k'}}-{\bf Q}_{k,k'}\right)},\\
    \label{eq: Z energy}
    \mathcal{Z}^\kappa_{E}&\!=\!\prod_{k=0}^{k_f}\int_{-\kappa}^{\kappa}d{w^{k}}e^{-\frac{{\bf w}{\bf Q}^{-1}{\bf w}}{2}}\,\,\,\text{s.t.}\,\,\,
    {\bf w}\!=\![w_0,\dots,w_{k_f}]\, ,
\end{align}
with the overlap matrix 
\begin{align}
    {\bf Q}_{k,k'}=\left\langle \frac{{\bf x}^{k}\cdot{\bf x}^{k'}}{N}\right\rangle
\end{align}
and $\langle\cdot\rangle$ the average over the measure from Eq.~(\ref{eq: connected measure}). Finally, the matrix $\hat{\bf Q}_{k,k'}$ is a set of fields that allows us to correctly set  these overlaps. By this we mean that by optimizing over these fields we have 
\begin{align}
       \left\langle \frac{{ x}^{k}\cdot{ x}^{k'}}{N}\right\rangle_{{\rm 1D},S}={\bf Q}_{k,k'}=\left\langle \frac{{\bf x}^{k}\cdot{\bf x}^{k'}}{N}\right\rangle
\end{align}
with $\langle\cdot\rangle_{{\rm 1D},S}$ the average over the measure from Eq.~(\ref{eq: Z entropy}).

The connected structure is ensured by imposing ${\bf Q}_{k,k\pm 1}=m$ in the overlap matrix. In other words, the free energy needs to be optimized over all overlaps except those involving nearest neighbors.
In \cite{barbier2025findingrightpathstatistical}, authors have analytically evaluated this free energy  by introducing a \emph{no-memory Ansatz}. Simply put, they assumed a Markovian geometry for the {connected} manifold:
\begin{align}
\label{eq: no-mem ansatz 1}
    \left[{\bf Q}^{-1}\right]_{k,k'}=\delta_{k',k\pm1}q_0+\delta_{k',k}q_1
\end{align}
and
\begin{align}
\label{eq: no-mem ansatz 2}
        \hat{\bf Q}_{k,k'}=\delta_{k',k\pm1}\hat{q}_0
\end{align}
such that
\begin{align}
\label{eq: no mem matrix correlation}
    {\bf Q}^{\rm no-mem.}_{k,k'}=m^{\vert k-k' \vert}\, .
\end{align}
It is called \emph{no-memory} because the memory kernel ${\bf Q}^{-1}$ only generates interactions between nearest-neighbor configurations ${\bf x}^k/{\bf x}^{k\pm 1}$. In the same paper, the authors demonstrated that this \emph{Ansatz} describes an existing {connected} manifold. However, it is not the one that optimizes the free energy; except in the trivial case of $\kappa\rightarrow+\infty$ or $\alpha\rightarrow 0$. Actually, they used such an \emph{Ansatz} because the saddle point of the free energy is difficult to estimate, especially in the limit $m \rightarrow 1$. First and foremost, taking $m \rightarrow 1$ is crucial, as it prevents the selected solutions from being \emph{isolated}; in this limit, they have infinitely-close neighbors. We recall that \emph{isolated} minima must be avoided, as they are algorithmically hard to sample. In this limit, the difficulty comes from the diverging size of the matrices ${\bf Q}$ and $\hat{\bf Q}$. To make this clearer, let us consider the overlap matrix in the \emph{no-memory} case -see Eq.~(\ref{eq: no mem matrix correlation})-. If we want the first and last configurations ${\bf x}^1/{\bf x}^{k_f}$ in our manifold to share an overlap $q=\mathcal{O}(1)$, we must take a number of steps that grow as $1/(1-m)$:
\begin{align}
    m^{k_f}=q\iff k_f=\frac{\log(q)}{\log(m)}\approx\frac{-\log{q}}{1-m}\, .
\end{align}
The same scaling applies to any correlation profile: as $m$ is sent to one, the size of the matrices diverges as $1/(1-m)$. To avoid this problem, one possible approach is to propose an \emph{Ansatz} for the matrices. In the following, we propose another approach. By coarse-graining the matrices, we will be able to approximate the saddle point for the free energy and optimize over a finite number of overlaps and fields (even when taking $m\rightarrow1$).

\section{Analysis of the connected free energy}
\label{sec: analysis}
\subsection{Coarse-graining approach}
Taking $m\rightarrow1$, we introduce a coarse-graining approach that simplifies the evaluation of the saddle point of the free energy. For this, we consider that a large number of entries in ${\bf Q}^{-1}$ and $\hat{\bf Q}$ are set to zero. More explicitly, we define a subgrid of variables $x$ and $w$
\begin{align}
    {\bf x}^*&=\left\{x^{*,k}\right\}_{k\in\left[\!\left[1,k_f^*\right]\!\right]}\,,\\
    {\bf w}^*&=\left\{w^{*,k}\right\}_{k\in\left[\!\left[1,k_f^*\right]\!\right]}\,,
\end{align}
for which the interactions are kept generic. The remaining configurations will follow the \emph{no-memory Ansatz} -see Eq.~(\ref{eq: no-mem ansatz 1},\ref{eq: no-mem ansatz 2})- to ensure the connected structure. For consistency, we will consider that between any two consecutive configurations $x^{*,k}$ and $x^{*,k+1}$ (respectively, $w^{*,k}$ and $w^{*,k+1}$) there are exactly $N_0$ \emph{no-memory} variables. This defines a coarse-grained overlap matrix ${\bf Q}^*$ such that
\begin{align}
{\bf Q}^*_{k,k'}=\left\langle \frac{{\bf x}^{k(N_0+1)}\cdot{\bf x}^{k'(N_0+1)}}{N}\right\rangle
\end{align}
with a paired coarse-grain field matrix $\hat{\bf Q}^*$.

\begin{figure}
    \centering
    \includegraphics[width=1\linewidth]{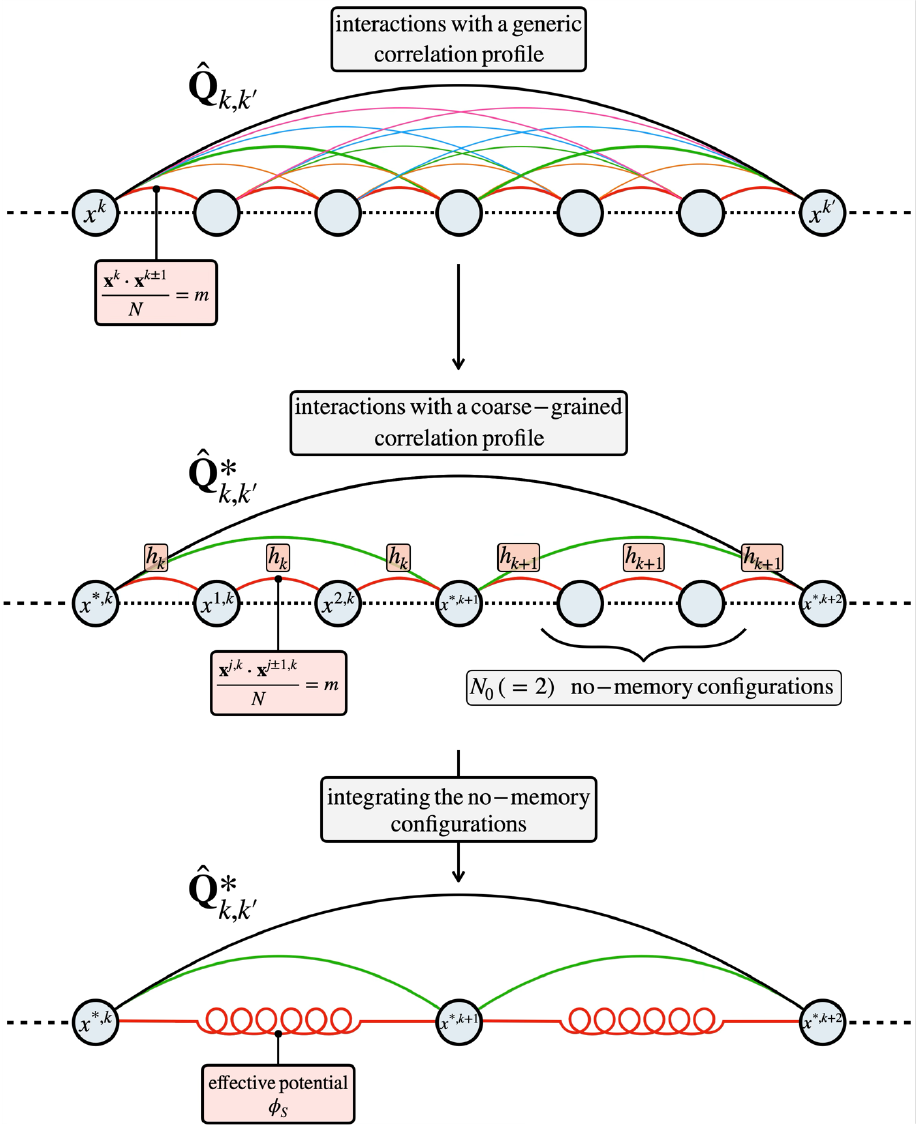}
    \caption{Illustration of the coarse-graining approach, focusing on the entropic partition function $\mathcal{Z}_{S}$ in the \emph{connected} free energy. At the top, we have the generic field matrix $\hat{\bf Q}$ that couples all the binary elements $\{x^k\}_{k\in[\![1,k_f]\!]}$ -represented with the colored lines-. The first step of the coarse-graining is to consider that only a subgrid of elements $\{x^{*,k}\}_{k\in[\![1,k_f^*]\!]}$ can have generic interactions -represented with the remaining black and green lines-. The remaining binary elements follow a \emph{no-memory} interactions profile -red lines- to ensure the connected structure. Because these elements only interacts with their nearest-neighbors, they can be integrated analytically. This second step generates a new effective potential $\phi_S$ between neighboring elements ${\bf x}^{*,k}/{\bf x}^{*,{k\pm 1}}$. The detailed computation for this coarse-graining approach can be found in App.~\ref{app: coarse-graining}.}
    \label{fig: 1}
\end{figure}

To better visualize this coarse-graining scheme, we provide an illustration in Fig.~\ref{fig: 1}. The advantage of this construction resides in the fact that \emph{no-memory} variables can be integrated out analytically, leaving only the ${\bf x}^*$ and ${\bf w}^*$ variables for the free energy optimization. Therefore, when we take the limit $m\rightarrow1$, this simply increases the number of \emph{no-memory} variables; ${\bf x}^*$ and ${\bf w}^*$ maintain their size. In practice, deriving the coarse-grained free energy is analytically tractable but computationally demanding; therefore, we detail the step-by-step computation in App.~\ref{app: coarse-graining}.

In the following, we will focus on characterizing the saddle point for the \emph{connected} free energy and its implications on the algorithmic hardness of the perceptron training.

\subsection{Characterization of the saddle point}
\label{sec: saddle-point charac}
\subsubsection{The correlation function}
\begin{figure}
    \centering
    $\kappa=0.75$
    \includegraphics[width=1\linewidth]{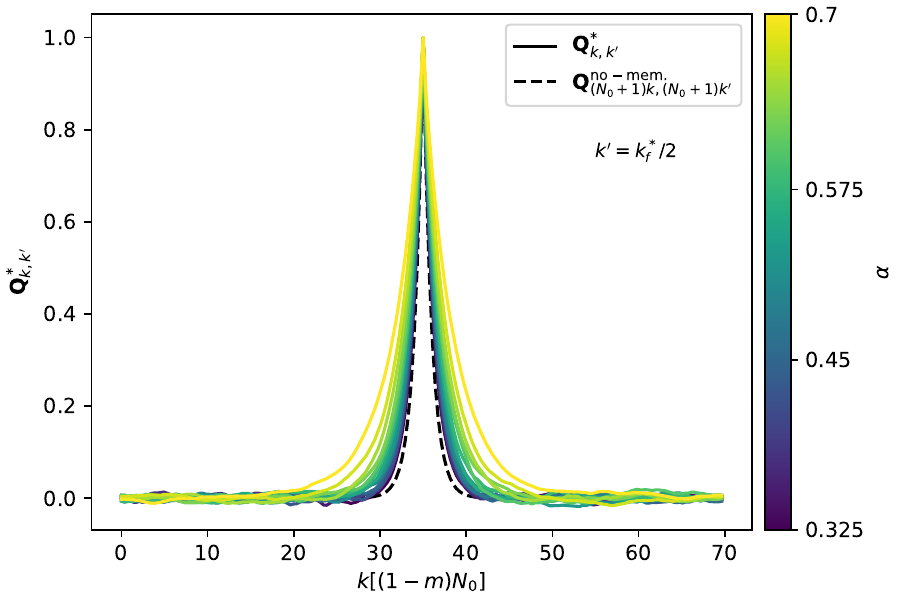}
    $\kappa=1.0$
    \includegraphics[width=1\linewidth]{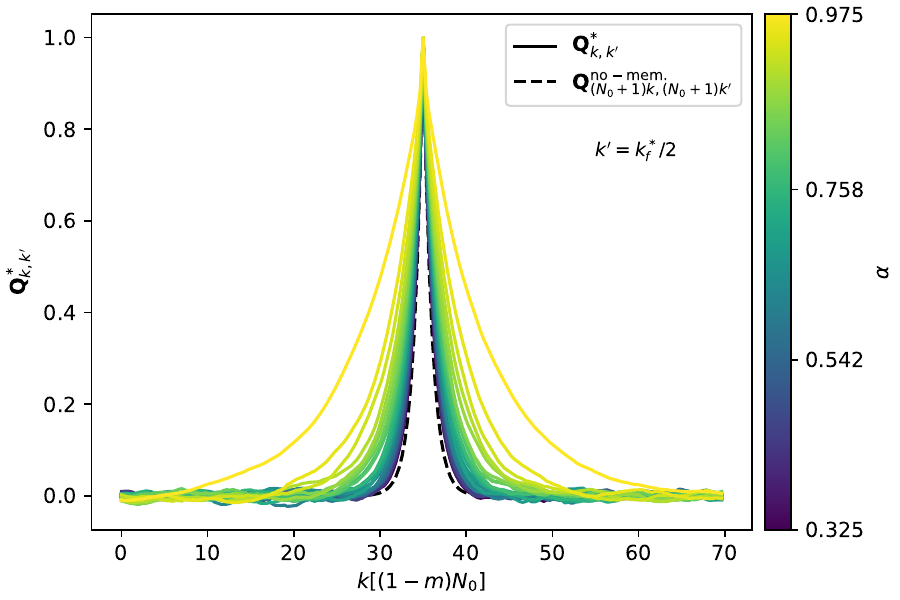}
    \caption{Evolution of the correlation profile ${\bf Q}^*_{k,k'}$ in the core of the connected manifold (i.e., $k'=k_f^*/2$). In both plots we increase $\alpha$ while keeping $\kappa$ fixed ($\kappa=\{0.75,1.0\}$). As a guide to the eye, the dashed profile corresponds to the \emph{no-memory} profile -see Eq.~(\ref{eq: correlation profile no-mem})-. The parameters we use for the coarse-graining are $N_0=700$, $m=0.9995$ and $k_f^*=200$. The maximum value of $\alpha$ we display (respectively $\alpha=0.7$ and $\alpha=0.975$) corresponds to the last value for which the saddle point of the \emph{connected} free energy exists.} 
    \label{fig: correlation linear}
\end{figure}

The main observable for characterizing the saddle point is the correlation function along paths
\begin{align}
    {\bf Q}^*_{k,k'}\!=\!\left\langle \!\frac{{\bf x}^{(N_0+1)k}\!\cdot\!{\bf x}^{(N_0+1)k'}}{N}\!\right\rangle\!=\!\left\langle\!\frac{x^{*,k}\!\cdot\! x^{*,k'}}{N}\!\right\rangle_{\rm \!\!1D,S}\!.
\end{align}
In Fig.~\ref{fig: correlation linear} we display the evolution of ${\bf Q}^*_{k,k'}$ (for two fixed values of $\kappa$) as $\alpha$ increases, i.e. as the number of classified patterns becomes greater.

Overall, the behavior of the correlation function remains the same as the difficulty of the task increases. Starting with a low $\alpha$ (or, equivalently, a high $\kappa$), training the SBP is almost trivial, and typical connected minima are nearly Markovian. Practically, this means that we have
\begin{align}
\label{eq: correlation profile no-mem}
   {\bf Q}^*_{k,k'}&\approx {\bf Q}^{\rm no-mem.}_{({N_0+1})k,({N_0+1})k'}\\
   &\approx m^{(N_0+1)\vert k-k' \vert}\nonumber\\
   &\hspace{-0.18cm}\underset{m\rightarrow1}{\approx}e^{-N_0(1-m)\vert k-k' \vert}\,.\nonumber
\end{align}
Here, we simply recover the fact that the \emph{no-memory Ansatz} gives the saddle point in the trivial case of $\alpha=0$.
As we increase the number of classified patterns, the typical connected manifold develops correlations between non-neighboring minima. In practical terms, the correlation function $\mathbf{Q}^*_{k,k'}$ along the paths exceeds that predicted by the \emph{no-memory Ansatz}. Finally, we identify a critical value of $\alpha$ (which depends on $\kappa$) above which the saddle point disappears.

Throughout the tuning of $\alpha$, four important features can be observed. First, for long distances, the overlap between two connected minima is zero (i.e., $\lim_{\vert k-k'\vert\rightarrow+\infty}{\bf Q}^{*}_{k,k'}=0$). This a posteriori validates the \emph{annealed} approximation we took to evaluate the \emph{connected} free energy ($\phi\approx\log[{\rm I\!E}_\xi(\mathcal{Z})]$). Indeed, we recall that this approximation is exact only if minima do not overlap \cite{aubin2019storage}, which is what we obtain here. The second feature we observe is that  
the correlation function follows an exponential trend, with a typical correlation length $\xi_k$ along the path. In other words, we have
\begin{align}
    {\bf Q}^*_{k,k'}&\approx e^{-\xi_kN_0(1-m)\vert k-k' \vert}\, .
\end{align}
In addition to this, the correlation function appears to maintain the same profile throughout the path. We find that the saddle point exhibits translation invariance (T.I.). This further simplifies the previous expression as
\begin{align}
    {\bf Q}^*_{k,k'}&\approx e^{-\xi_kN_0(1-m)\vert k-k' \vert}\\
    &\hspace{-0.08cm}\underset{\rm T.I.}{\approx }e^{-\xi N_0(1-m)\vert k-k' \vert}\, . \nonumber
\end{align}
Lastly, we note that this correlation length  diverges rapidly close to the phase transition; i.e., where the saddle point disappears. 

\begin{figure}
    \centering
    \includegraphics[width=1\linewidth]{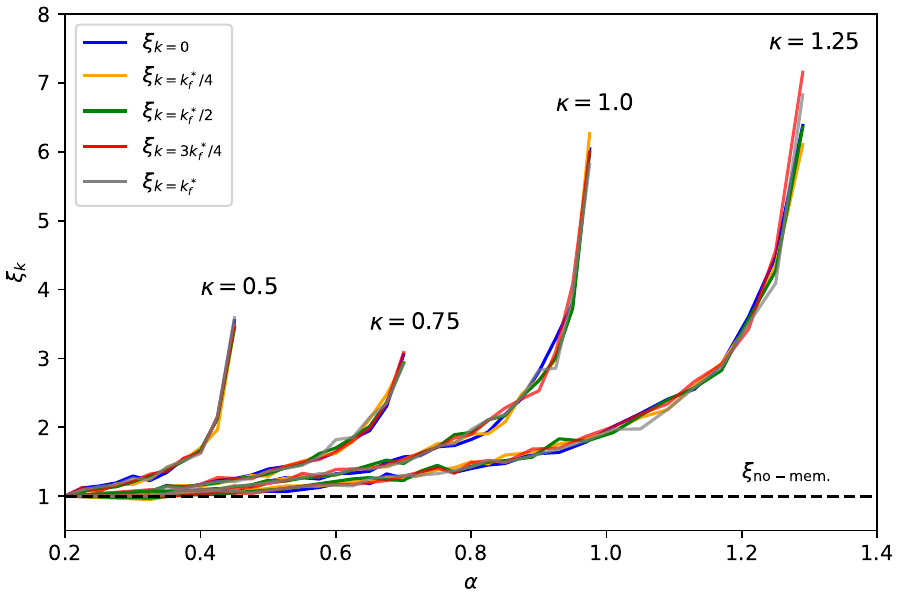}
    \caption{Evolution of the correlation length $\xi_k$ as a function of $\alpha$. Four threshold values are displayed: $\kappa\in\{0.5,0.75,1.0,1.25\}$. More particularly, we plot  $\xi_k$ for five positions in the path: at the edges ($k=\{0,k_f^*\}$), in the middle ($k=k_f^*/2$), and in the first and last quarters ($k=\{k_f^*/4,3k_f^*/4\}$). In dashed black, we indicate the correlation length for \emph{no-memory} minima. This plot highlights the translation invariance in the correlation profile, as $\xi_k$ appears to be independent of the position $k$. It also illustrates the divergence of the correlation length at the connected transition (i.e, for $\alpha=\alpha_{\rm connected}$). When $\alpha>\alpha_{\rm connected}$, the free energy does not admit any saddle point.}  
    \label{fig: correlation length}
\end{figure}

To better visualize these features, we first plot in Fig.~\ref{fig: correlation length} the correlation length $\xi_k$ along the path ($k\in[\![1,k_f^*]\!]$). As seen in the figure, the value of $\xi_k$ does not depend on the position $k$ for which it is evaluated. We can also observe its divergence as we increase $\alpha$ and approach the phase transition. For $\kappa\in\{0.75,1.0,1.25\}$, we observe a clear power-law trend: $\xi\sim[\alpha_{\rm connected}(\kappa)-\alpha]^{-\lambda(\kappa)}$, where $\alpha_{\rm connected}(\kappa) = \{0.70, 0.98, 1.33\}$ and $\lambda(\kappa) = \{0.23, 0.38, 0.60\}$, respectively. For lower values of $\kappa$, tracking the saddle point as a function of $\alpha$ (especially near the transition) is difficult, since the correlation length diverges rapidly. As a consequence, the power-law trend is less clear, and estimating the exponents becomes unreliable.

\begin{figure}
    \centering
    $\kappa=0.75$
    \includegraphics[width=1\linewidth]{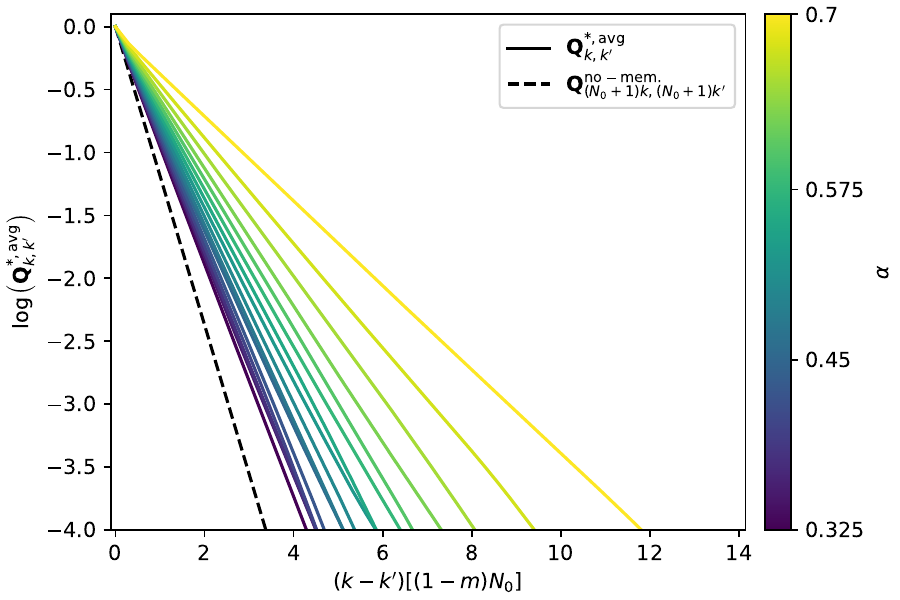}
    $\kappa=1.0$
    \includegraphics[width=1\linewidth]{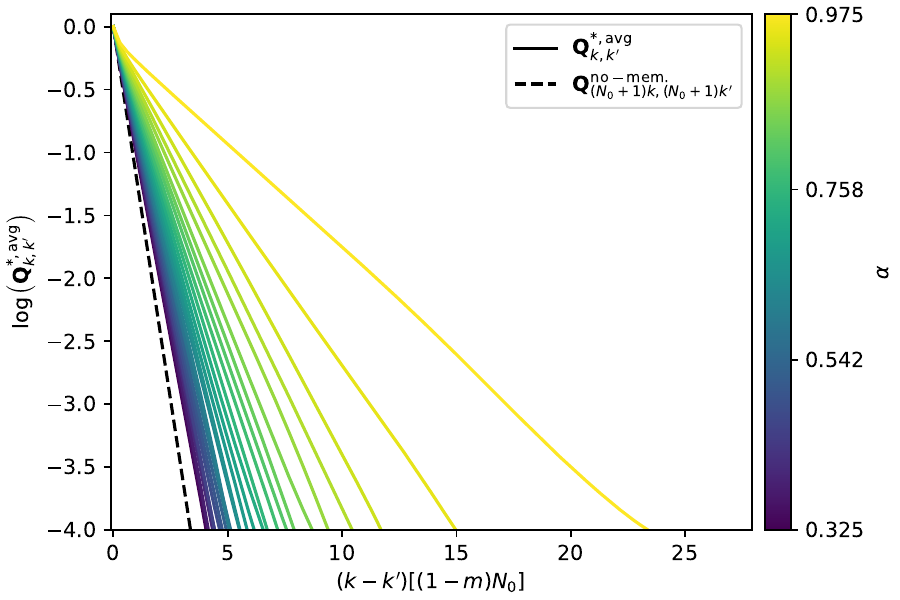}
    \caption{Evolution of the averaged correlation profile ${\bf Q}^{*,{\rm avg}}_{k,k'}$ (averaged along the path) as the classification task becomes more difficult. In this case, we increase $\alpha$ while keeping $\kappa$ fixed ($\kappa=\{0.75,1.0\}$). To highlight the exponential trend, the correlation function is plotted in log-linear scale. With this scaling, the slope of the curves corresponds to the correlation length $\xi$. The parameters for the coarse-graining are $N_0=700$, $m=0.9995$ and $k_f^*=200$. The dashed line corresponds to the \emph{no-memory} profile.}
    \label{fig: correlation log}
\end{figure}

To further illustrate the exponential trend in the correlation function, we plot in Fig.~\ref{fig: correlation log} the averaged correlation function $\bf Q^{*,\rm avg}_{k,k'}$ (averaged along the path) in log-linear scale. This allows us to clearly identify the linear slope, indicative of exponential decay in the correlation function. It also illustrates how rapidly the profile changes near the transition. Taking the case $\kappa=1.0$ for example, the  correlation function barely deviates from the no-memory prediction for a large range of $\alpha$ ($\alpha\in[0.325,0.8]$). Only when we get close to the transition ($\alpha_{\rm connected}\approx0.98$) does the slope suddenly increase.


\subsection{Margins distribution}
In this section, we will focus on another key observable that characterizes the saddle point: the margins distribution. More explicitly, we will study the behavior of 
\begin{align}
    P^k\left(w\right)=\left\langle \delta\left(w\!-\!w^k\right)\right\rangle_{{\rm 1D},E}= \left\langle\! \delta\!\left(w\!-\!\frac{\xi^\mu\cdot{\bf x}^k}{\sqrt{N}}\right)\!\right\rangle
    \end{align}
irrespective of the value of $\mu~(\in[\![1,M ]\!])$. The average $\langle \cdot\rangle_{{\rm 1D},E}$ corresponds to the margin measure introduced in Eq.~(\ref{eq: Z energy}). 

This observable is of prime interest, as it
characterizes the outputs returned by the optimized network (for the input patterns used in training). Importantly, with it, we can assess how well a configuration ${\bf x}^k$ performs on the perceptron classification task. For example, a minimum is more robust if most of its outputs (or margins) lie away from the decision boundaries $w = \pm \kappa$. Indeed, if input patterns are perturbed, a robust minimum should have as few misclassified inputs as possible, and hence maintains its margins away from the decision boundaries. As an example, the typical solutions of the SBP have a truncated Gaussian margin distribution
\begin{align}
    P^{\rm typ.}(w)\propto \Theta(\kappa-\vert w\vert)e^{-w^2/2}\,.
\end{align}
With this distribution, typical minima accumulate margins close to $ w=\pm\kappa$, since $P^{\rm typ.}(w=\pm \kappa)\neq0$. This makes them non-robust, and also \emph{isolated} \cite{krauth89storage,Barbier2025}.

\begin{figure}
    \centering
    $\kappa=1\,, ~\alpha=0.4$
    \includegraphics[width=1\linewidth]{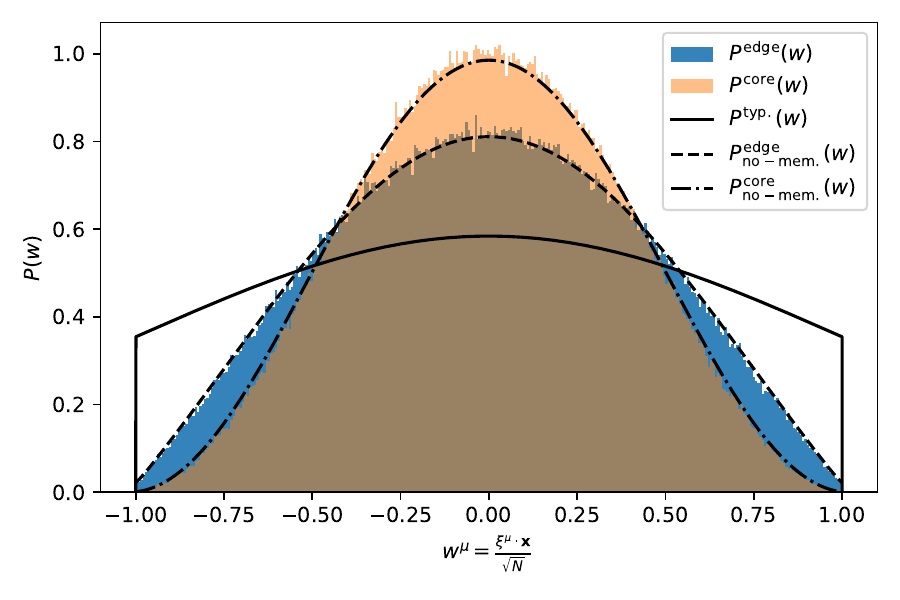}
    $\kappa=1\,, ~\alpha=0.85$
    \includegraphics[width=1\linewidth]{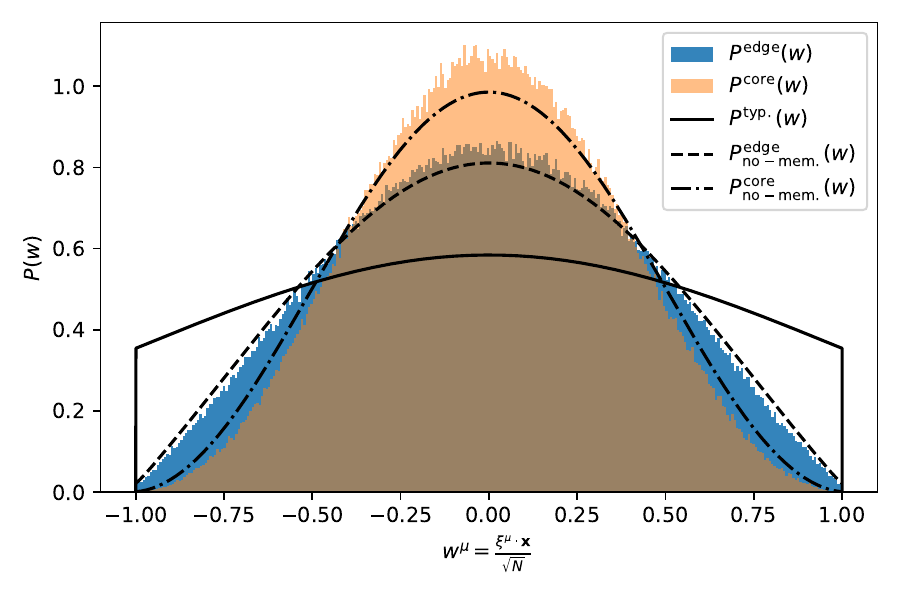}
    $\kappa=1\,, ~\alpha=0.975$
    \includegraphics[width=1\linewidth]{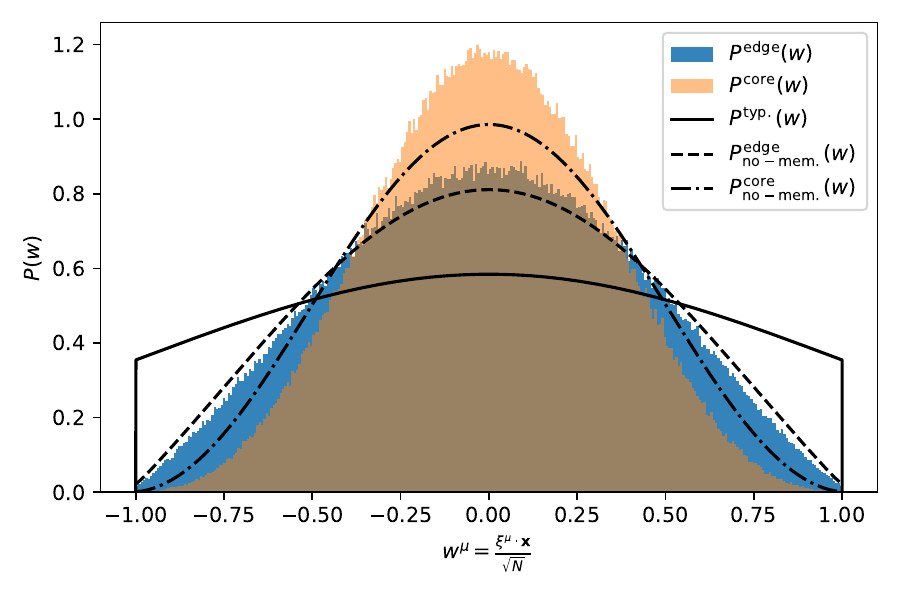}
    \caption{Margin distribution for the typical connected minima. For each case ($\alpha=\{0.4,0.85,0.975\}$ with fixed $\kappa=1.0$), we make the distinction between minima at the edges of the manifold (in blue) and minima in the core (in orange). In more detail, the edge distribution is obtained by evaluating $\langle \delta(w\!-\!w^{k=0})\rangle_{{\rm 1D},E}$ and the core distribution by evaluating $\langle \delta(w\!-\!w^{k=k_f^*/2})\rangle_{{\rm 1D},E}$; where $\langle \cdot\rangle_{{\rm 1D},E}$ corresponds to the margin measure introduced in Eq.~(\ref{eq: Z energy}). As a guide to the eye, we also display the margin distribution for typical minima (full line) and for minima at the edges (dashed line) and in the core (dash-dotted line) of the \emph{no-memory} manifold
 \cite{barbier2025findingrightpathstatistical}.}
    \label{fig: margins distribution}
\end{figure}

In Fig.~\ref{fig: margins distribution}, we plot the margin distributions for typical connected minima. At the top, we display a low $\alpha$ case ($\kappa=1$ and $\alpha=0.4$), for which the correlation function is close to the \emph{no-memory} profile. At the bottom, $\alpha$ is higher (close to the transition at $\alpha\approx0.98$ and $\kappa=1$), and typical connected solutions have a correlation length significantly greater than that of \emph{no-memory} minima. As already mentioned in \cite{barbier2025findingrightpathstatistical}, two cases need to be distinguished:
\begin{itemize}
    \item In blue, minima at the edges of the \emph{connected} manifold (${\bf x}^{k\approx0}$ or ${\bf x}^{k\approx k_f}$).
    \item In orange, minima in the core of the manifold (${\bf x}^{k\approx k_f/2}$).
\end{itemize}
Regardless of $\alpha$ and the position in the connected manifold, the margin density is always low close to the decision boundaries $w=\pm\kappa$. This shows the tight link between connectivity and robustness. To be connected, the selected solutions have to be significantly more robust than typical minima (and vice-versa). Then, we can also note that solutions at the edges are less robust than those at the core. It is again the interplay between robustness and connectivity. In the core, a solution ${\bf x}^{k\approx k_f/2}$ sees a long path of connected minima on its left ${\bf x}^{k'(<k)}$ and on its right ${\bf x}^{k'(>k)}$. At the edges, minima ${\bf x}^{k\approx0}$ only see a path on their right ${\bf x}^{k'(>k)}$. Thus, when in the core, the effect of connectivity is ``doubled'', and minima are more robust.

If we now compare the margin distributions for different values of $\alpha$, typical connected minima become increasingly robust as the classification task becomes more difficult (i.e., as $\alpha$ increases). This means that the minima accessible for the largest range of parameters $\alpha/\kappa$ are also the most robust. Even more interestingly, putting this figure in parallel with Fig.~\ref{fig: correlation length}, we see that robustness increases with the typical correlation length $\xi$. For $\alpha=0.4$, $\xi$ is close to the correlation length of \emph{no-memory} minima, and the margin distributions (both at the edges and in the core) match the predictions in \cite{barbier2025findingrightpathstatistical} based on \emph{no-memory Ansatz}. For $\alpha=0.975$, the correlation length is significantly greater, and the distributions (both at the edges and in the core) are more compact. In fact, this trend has also been observed with protein models, where proteins following evolutionary paths with greater correlation length are significantly more robust \cite{Mauri2023,Greenbury2022,Wu2016}. Thus, this interplay between correlation length and robustness appears to be a universal property of connected regions in rugged landscapes.

Finally, Fig.~\ref{fig: margins distribution} allows us to understand one of the main features observed in \cite{barbier2025findingrightpathstatistical}. In this paper, the authors used a Monte Carlo algorithm to access the edge minima of the \emph{no-memory} manifold. To function, this algorithm relied on a specific effective loss (specifically tuned to sample the margin distribution of \emph{no-memory} minima at the edges of cluster). They noted is that even the slightest discrepancy in this effective loss (hence in the margin distribution they targeted) visibly changes the algorithmic transition. If we focus on the margin distribution for the \emph{edges} in Fig.~\ref{fig: margins distribution}, we can understand the origin of this phenomenon. In fact, we can note that the typical margin distribution never deviates significantly from the \emph{no-memory} distribution (regardless of $\alpha$). In other words, typical connected minima and \emph{no-memory} ones  are hardly distinguishable (at the edges) based on their margin distributions. Thus, it is very easy to target the wrong minima by making a slight numerical error when evaluating the margin distribution; which is used to derive the effective loss implemented in algorithms. Now for $\kappa=1.0$, we have a $25\%$ difference between the algorithmic transition for \emph{no-memory} minima and typical connected minima  ($\alpha^{\rm no-mem}_{\rm loc. \, stab.}\approx0.74$ and $\alpha_{\rm connected}\approx0.98$). Hence, targeting the wrong minima (with just a slight numerical mistake in the effective loss) can significantly shift the observed algorithmic transition.

\subsection{Phase diagram}

In this section, we summarize the key features related to the algorithmic hardness of SBP training.

\begin{figure}
    \centering
    \includegraphics[width=1\linewidth]{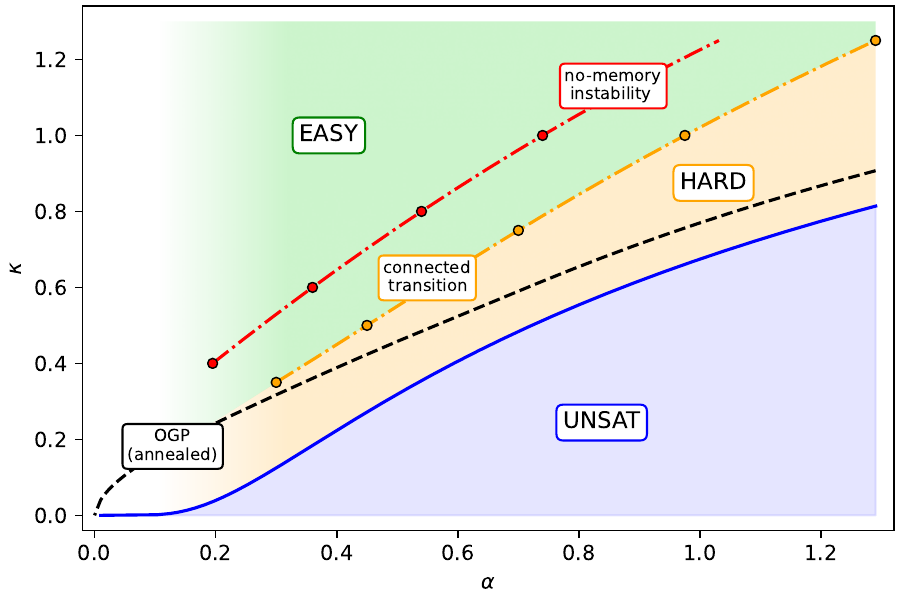}
    \caption{Phase diagram of the SBP training. The green area represent the easy phase, where connected minima exist and are accessible by local algorithms. The intermediate red dashed line corresponds to the limit for which \emph{no-memory} minima are stable. Below this line, \emph{no-memory} minima cannot be sampled \cite{barbier2025findingrightpathstatistical}. Below the connected transition (in orange), connected minima do not exist and training the SBP network becomes hard. Both the \emph{no-memory} and \emph{connected} transitions are evaluated for several values of $\kappa$ (highlighted with dots). The colored dashed-dot lines are degree-3 polynomial interpolations. Finally, in blue, we have the UNSAT phase; where solving the SBP classification task is impossible. We highlighted with a dashed black line the OGP transition \cite{barbier2023atypical}. It represented the line above which there exist minima which are not \emph{isolated}. }
    \label{fig: phase diagram}
\end{figure}

To begin with, we detail in Fig.~\ref{fig: phase diagram} the different regimes we obtained for training. When $\kappa$ is large (respectively, $\alpha$ is low), training the network is easy. The decision boundaries are wide enough (respectively, the number of input patterns is low enough) to find connected solutions in the loss landscape. These connected solutions are accessible and, as shown in \cite{barbier2025findingrightpathstatistical}, can be sampled using a modified Monte Carlo algorithm. In particular, above the red line, \emph{no-memory} minima that exist and can be sampled. This easy training regime stops at the orange line, where the \emph{connected} free energy loses its saddle point. For lower $\kappa$ (respectively, higher $\alpha$), training the SBP becomes hard. In other words, in the orange area, there are still binary configurations that solve the classification task, but they are \emph{isolated} \cite{krauth89storage,Barbier2025}. Surrounded by large high-loss barriers, these minima are difficult to access with local algorithms. Along with isolation, these minima also lack robustness, making them even more inadequate for realistic classification.

If we keep decreasing $\kappa$ (or increasing $\alpha$), we cross the UNSAT line (in blue) and the classification task becomes unsolvable. This means that, in the area highlighted in blue, there are no solutions to the SBP problem. Finally, to situate our approach in relation to other state-of-the-art methods, we have added to this phase diagram the overlap-gap transition developed in \cite{barbier2023atypical}. In simple terms, this approach consists of computing the value of $\kappa$ for which the most robust solutions (i.e., the configuration solving the task for $\kappa_0 = \kappa_{\rm SAT}$ and fixed $\alpha$) are no longer \emph{isolated} ($\kappa > \kappa_0$). In this case, being \emph{not isolated} means being able to find other minima at any distance from those ``most robust'' solutions. As we can see, this overlap-gap transition is close to the connected one for low $\alpha/\kappa$, and deviates from it for larger values of $\alpha/\kappa$. It also predicts a greater range for the easy training regime. This discrepancy is due to the fact that connectivity is a more restrictive criterion than being \emph{not isolated}. Indeed, a solution can have other minima at all distances away from it; this does not mean that paths connecting them exist. When neighboring minima exist but low-loss paths cannot be found, local algorithms simply cannot exploit these regions for training. This is the reason why the OGP criterion overshoots the range of the easy training regime.

\begin{figure*}[t]
    \centering
    \includegraphics[width=\textwidth]{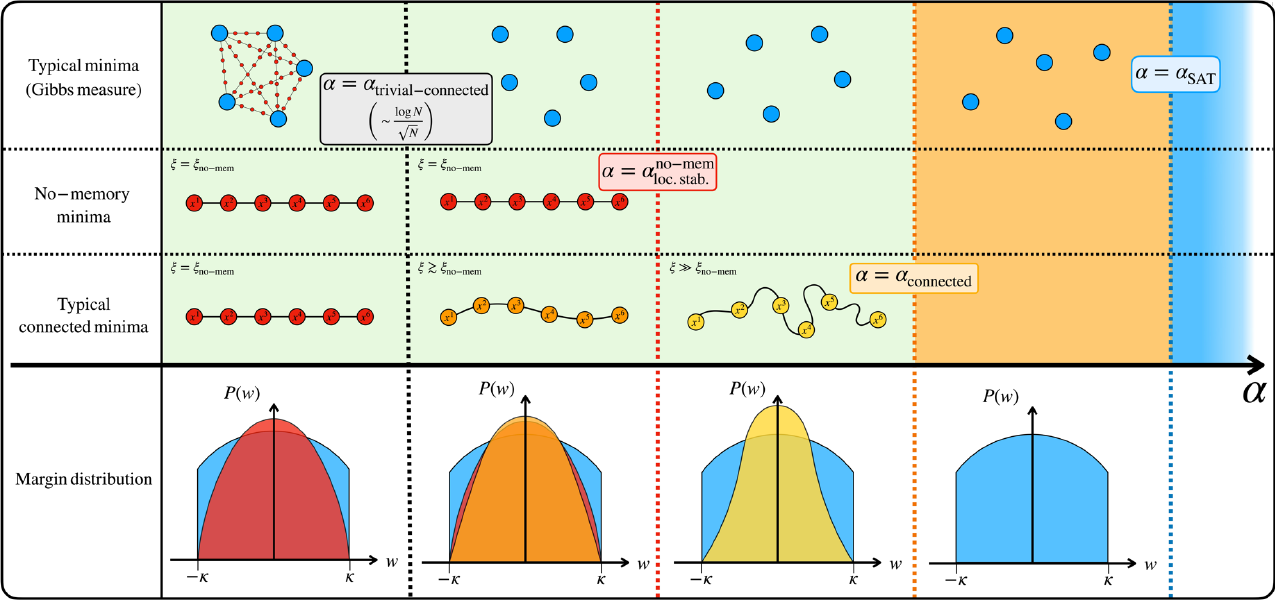}
    \caption{Table summarizing the main features of the SBP minima. We have separated the cases between typical, \emph{no-memory} and typical connected minima. With $\kappa$ fixed, the training becomes more difficult as $\alpha$ increases. When it is very small, i.e. $\alpha\sim \log(N)/\sqrt{N}$, the training is trivial as even the typical minima (sampled from the standard Gibbs measure) are connected \cite{Barbier2025}. As $\alpha$ increases, the first transition we observe is the instability of \emph{no-memory} minima. Beyond the red dashed line, \emph{no-memory} minima cannot be sampled with local algorithms. Then, connected minima continue to exist up to $\alpha=\alpha_{\rm connected}$. As the task become more difficult (and while training remains easy), the correlation length $\xi$ for these minima increases. For $\alpha>\alpha_{\rm connected}$, the loss landscape do not contain any connected minima, training the network becomes hard. Finally, for $\alpha>\alpha_{\rm SAT}$, the SBP task has no solutions. We illustrated the margin distributions for each type of minima with color-coded drawings, without distinguishing between edge and core minima.}
    \label{fig: summary}
\end{figure*}

Finally, in Fig.~\ref{fig: summary}, we detail the behavior of the typical, \emph{no-memory} and typical connected minima as the classification task becomes more and more difficult. First, when the task is extremely simple (either $\alpha\sim \log(N)/\sqrt{N}$ with $\kappa\sim \mathcal{O}(1)$, or $\kappa\sim \sqrt{\log(N)}$ with $\alpha\sim \mathcal{O}(1)$), all minima are connected with high probability \cite{Barbier2025}. In this regime, even typical minima (which dominate the Gibbs measure \cite{krauth89storage}) can be accessed efficiently with local algorithms. In fact, a simple Monte Carlo routine (with no particular modifications) can train the network \cite{Barbier2025}. As $\alpha$ increases, only \emph{no-memory} and typical connected solutions remain accessible. Because of their greater correlation length, typical connected solutions are more robust than the \emph{no-memory} ones. For $\alpha>\alpha_{\rm loc.\, stab.}^{no-mem}$, only the typical connected minima remain accessible. Their correlation length continues to increase, leading to a greater robustness. When $\alpha>\alpha_{\rm connected}$, there are no more connected minima and training becomes hard. Finally, for $\alpha>\alpha_{\rm SAT}$, the SBP classification task has no solutions.

\section{Conclusion}
To conclude this article, let us recap our investigation of the \emph{connected} free energy and its saddle point. Here, our main objective was to identify the $\alpha/\kappa$ region for which the SBP admits connected minima and therefore can be efficiently trained. A secondary aim was to characterize these connected regions, in particular with regard to robustness.

First, to compute the saddle point in the limit $m\rightarrow1$, we introduced a coarse-graining approach to limit the difficulty of the free energy optimization. Then, by analyzing the saddle point, we observed that connected minima exist only for values of $\alpha$ that are low enough (respectively, values of $\kappa$ that are large enough). This existence defines the ``easy training" regime for the SBP, because these minima can be efficiently probed by local algorithms. In this regime, the correlation profile along a connected path follows an exponential decay (with a typical correlation length $\xi$). As the task becomes harder, the correlation length increases and finally diverges at the transition $\alpha=\alpha_{\rm connected}$ (for fixed $\kappa$). 

In terms of margin distribution, we observe the same manifold structure as in \cite{barbier2025findingrightpathstatistical}: robust minima at the edges (${\bf x}^{k\approx 0}$ or ${\bf x}^{k\approx k_f}$) and even more robust minima in the core (${\bf x}^{k\approx k_f/2}$). As the classification task becomes more difficult (and the correlation length along paths increases), typical connected minima become more robust. This interplay between robustness and correlation length appears to be a universal feature of connected manifolds; it has already been observed in the mutational paths of proteins \cite{Mauri2023,Wu2016}. Finally, the discrepancy between the margin distributions of $\emph{no-memory}$ minima and typical connected minima is visible but remains relatively small. This explains why accurately identifying the algorithmic transition for $\emph{no-memory}$ minima can be difficult \cite{barbier2025findingrightpathstatistical}: the slightest numerical mistake in the margin distribution being targeted can significantly shift the transition point.

The SBP is an extremely rich toy model, particularly for understanding the behavior of rugged landscapes. It highlights the need to introduce new statistical mechanics tools to properly understand algorithmic transitions. This article shows why the \emph{connected} ensemble appears as a credible alternative to the standard Gibbs measure for characterizing landscapes. Its main strengths are based on a key advantage: being able to identify regions where minima aggregate and form paths. Because we can derive and analyze the statistics of these paths, designing local algorithms that will access these regions becomes extremely easy \cite{barbier2025findingrightpathstatistical}. 

Several open questions remain. For example, can we exploit other atypical connected regions by tuning the Lagrange multipliers $\{y_k\}_{k\in[\![1,k_f]\!]}$ away from one -see Eq.~(\ref{eq: connected measure})-? What happens to the \emph{connected} free energy when the \emph{annealed} approximation is not valid? Can we exploit the \emph{connected} ensemble to study more realistic models and derive efficient algorithms? All these questions will be the focus of further studies, revolving around this new statistical mechanics approach.

\begin{acknowledgments}
We would like to thank Riccardo Zecchina, Jérôme Garnier-Brun, Luca Saglietti, Gianmarco Perrupato, and Enrico Malatesta for the many discussions and guidance about this project.
\end{acknowledgments}

\appendix

\section{Computing of the generic free energy}
\label{app: General comp}
In this section we detail how to compute the $quenched$ free energy
\begin{align}
    \phi={\rm I\!E}_{\xi}[\log(\mathcal{Z})]\underset{y_0\rightarrow 0}{=}\frac{{\rm I\!E}_{\xi}[\mathcal{Z}^{y_0}]-1}{y_0}\
\end{align}
in the chain case ($y_k=1$ for $k>0$). In this regime we have
\begin{align}
\hspace{-0.22cm}{\rm I\!E}_\xi[\mathcal{Z}^{y_0}]={\rm I\!E}_{\xi}\left\{\prod_{a=1}^{y_0}\prod_{k=0}^{k_f}\sum_{{\bf x}^{k,a}\in\Sigma^{N,m}_{{\bf x}^{k-1,a}}}e^{\mathcal{L}_{\rm SBP}\left({\bf x}^{{k,a}}\right)}\!\right\}\,.
\end{align}
So as to perform the average over the patterns distribution we will introduce the margins as
\begin{align}
    e^{\mathcal{L}_{\rm SBP}\left({\bf x}^{k,a}\right)}&\!=\!\!\prod_{\mu=1}^M\!\Theta\left(\kappa-\left\vert\frac{\xi^\mu\cdot{\bf x}^{k,a}}{\sqrt{N}}\right\vert\right)\\
    &\hspace{-1.5cm}=\!\!\prod_{\mu=1}^M\int_{-\kappa}^{+\kappa} \!\!dw^{k,a,\mu}\!\int_{-\infty}^{+\infty}\!\!d\hat{w}^{k,a,\mu}e^{i\hat{w}^{k,a,\mu}\left(\!w^{k,a,\mu}-\frac{\xi^\mu\cdot{\bf x}^{k,a}}{\sqrt{N}}\!\right)}\!.\nonumber
\end{align}
The successive integration over the patterns $\{\xi^\mu\}_{\mu\in{[\![1,M]\!]}}$ distribution and after over the fields $\{\hat{w}^{{\bf P}_k,\mu}\}_{\mu\in{[\![1,M]\!]}}$ is trivial as it simply involves Gaussian integrals. It yields
\begin{align}
    {\rm I\!E}_\xi[\mathcal{Z}^{y_0}]=&\\
    &\hspace{-1.5cm}\prod_{a=1}^{y_0}\prod_{k=0}^{k_f}\prod_{\mu=1}^M\sum_{{\bf x}^{{k,a}}\in\Sigma^{N,m}_{{\bf x}^{k-1,a}}}\int_{-\kappa}^{\kappa}\!\!dw^{k,a,\mu}\frac{e^{-\frac{{\bf w}\underline{\bf Q_{\bf x}}^{-1}{\bf w}}{2}}}{\mathcal{N}_{\bf x}}\nonumber
\end{align}
with
\begin{align}
   \mathcal{N}_{\bf x}=\prod_{a=1}^{y_0}\prod_{k=0}^{k_f}\prod_{\mu=1}^M\int_{-\infty}^{+\infty} \!\!dw^{k,a,\mu}{e^{-\frac{{\bf w}\underline{\bf Q_{\bf x}}^{-1}{\bf w}}{2}}}\nonumber
\end{align}
and
\begin{align}
    \underline{\bf Q_{\bf x}}_{k,a,{k'},b}^{\mu,\mu'}=\delta_{\mu,\mu'}\,\frac{{\bf x}^{k,a}\cdot{\bf x}^{{k',b}}}{N}\, .
\end{align}
Another way to obtain this result is to recognize that (without the loss) the margins are random Gaussian variables, following the means and correlations
\begin{align}
\label{eq: mean margin}
    {\rm I\!E}_{\xi}\left[w^{k,a,\mu}\right]&={\rm I\!E}_{\xi}\left[\frac{\xi^\mu\cdot{\bf x}^{k,a}}{\sqrt{N}}\right]\\&=0\, ,\nonumber\\
\label{eq: correlation margin}
    {\rm I\!E}_{\xi}\left[w^{k,a,\mu}w^{k',b,\mu'}\right]&={\rm I\!E}_{\xi}\left[\frac{(\xi^\mu\cdot{\bf x}^{k,a})(\xi^{\mu'}\cdot{\bf x}^{{k',b}})}{{N}}\right]\\&=\delta_{\mu,\mu'}\,\frac{{\bf x}^{k,a}\cdot{\bf x}^{{k',b}}}{N}\, ,\nonumber
\end{align}
and to simply introduce this distribution in the partition function. The last step is to decouple the margins contribution from the hypercube summation. For this we introduce magnetic fields as
\begin{align}
    &\prod_{a=1}^{y_0}\prod_{k=0}^{k_f}\sum_{{\bf x}^{{k,a}}\in\Sigma^{N,m}_{{\bf x}^{k-1,a}}}f\left(\{{\bf x}^{k,a}\cdot{\bf x}^{{k',b}}\}\right)=\\
    &\int\prod_{a,b=1}^{y_0} \prod_{k,k'=0}^{k_f}d\hat{\bf Q}_{k,a,k',b}d{\bf Q}_{k,a,k',b}f\left(\{{\bf Q}_{k,a,k',b}\}\right)\nonumber\\
    &\hspace{-0.3cm}\times\! \prod_{a=1}^{y_0}\prod_{k=0}^{k_f}\sum_{{\bf x}^{{ k,a}}\in\Sigma^{N}}e^{\underset{a,b}{\sum}\,\underset{k,{k'}}{\sum} i\hat{\bf Q}_{k,a,k',b}\left[{\bf x}^{k,a}\cdot{\bf x}^{{k',b}}-N{\bf Q}_{k,a,k',b}\right]}\nonumber\,.
\end{align}
Before writing this decoupling for the SBP model, we can also mention that the previous expression can be evaluated with a saddle point approximation when $N\rightarrow+\infty$:
\begin{align}
     &\lim_{N\rightarrow+\infty}\prod_{a=1}^{y_0}\prod_{k=0}^{k_f}\sum_{{\bf x}^{{k,a}}\in\Sigma^{N,m}_{{\bf x}^{k-1,a}}}f\left(\{{\bf x}^{k,a}\cdot{\bf x}^{{k',b}}\}\right)=\nonumber\\
    & \hspace{0cm}\underset{
    \tiny\begin{array}{c}
         \hat{\bf Q}_{k,a,k',b}  \\
         {\bf Q}_{k,a,k',b}\\
           ({\bf Q}_{k,a,k\pm1,a}\!=\!m)
    \end{array}}{\rm opt}\left\{ \vphantom{\prod_{\sum_{{\bf x}^{{\bf j}_{ k}}\!\in\Sigma^{N}}{\bf j}_{k}}}f\left(\{{\bf Q}_{k,a,k',b}\}\right)\right.\\
    &\hspace{0cm}\times\prod_{a=1}^{y_0}\prod_{k=0}^{k_f} \sum_{{\bf x}^{{ k,a}}\in\Sigma^{N}}\left.\!e^{\underset{a,b}{\sum}\,\underset{k,{k'}}{\sum}\!\hat{\bf Q}_{k,a,k',b}\!\left[{\bf x}^{k,a}\cdot{\bf x}^{{k',b}}-N{\bf Q}_{k,a,k',b}\right]}\!\right\}\!.\nonumber
\end{align}
Applying this formula to our setup we obtain
\begin{align}
    {\rm I\!E}_\xi[\mathcal{Z}^{y_0}]=&\underset{
    \tiny\begin{array}{c}
         \hat{\bf Q}_{k,a,k',b}  \\
         {\bf Q}_{k,a,k',b}\\
           ({\bf Q}_{k,a,k\pm1,a}\!=\!m)
    \end{array}}{\rm opt}\left\{\tilde{\mathcal Z} \right\}
\end{align}
with
\begin{align}
    \tilde{\mathcal Z}=&\prod_{a=1}^{y_0}\prod_{k=0}^{k_f}\prod_{\mu=1}^M\int_{-\kappa}^{\kappa}{dw^{k,a,\mu}}\frac{e^{-\frac{{\bf w}\underline{\bf Q}^{-1}{\bf w}}{2}}}{\mathcal{N}}\\
    &\times\prod_{a=1}^{y_0}\prod_{k=0}^{k_f}\sum_{{\bf x}^{{ k,a}}\in\Sigma^{N}}\nonumber\\
    &\hspace{0.cm}\times e^{\underset{a,b}{\sum}\,\underset{k,{k'}}{\sum}\!\hat{\bf Q}_{k,a,k',b}\!\left[{\bf x}^{k,a}\cdot{\bf x}^{{k',b}}-N{\bf Q}_{k,a,k',b}\right]}\nonumber\! ,
\end{align}
the normalization $\mathcal{N}$ being
\begin{align}
\mathcal{N}=\prod_{a=1}^{y_0}\prod_{k=0}^{k_f}\prod_{\mu=1}^M\int_{-\infty}^{+\infty}d{w^{k,a,\mu}} e^{-\frac{{\bf w}\underline{\bf Q}^{-1}{\bf w}}{2}}
\end{align}
and $\underline{\bf Q}^{-1}$ being the inverse of the overlap matrix
\begin{align}
     \underline{\bf Q}_{k,a,{k'},b}^{\mu,\mu'}=\delta_{\mu,\mu'}\,  {\bf Q}_{k,a,{k'},b}\, .
\end{align}
Finally, we can recognize that all directions $i\in[\![1,N]\!]$ of the hypercube $\Sigma^N$ are decoupled -same for the margins directions $\mu\in[\![1,M]\!]$-, which yields
\begin{align}
    \tilde{\mathcal Z}=&\left\{\prod_{a=1}^{y_0}\prod_{k=0}^{k_f}\int_{-\kappa}^{\kappa}d{w^{k,a}}\frac{e^{-\frac{{\bf w}{\bf Q}^{-1}{\bf w}}{2}}}{\mathcal{N}}\right\}^M\\
    &\hspace{-0.8cm}\times\!\!\left\{\!\prod_{a=1}^{y_0}\prod_{k=0}^{k_f}\sum_{{x}^{{k,a}}=\pm 1}e^{\underset{a,b}{\sum}\,\underset{k,{k'}}{\sum}\!\hat{\bf Q}_{k,a,k',b}\!\left[{x}^{{k,a}}{ x}^{{k',b}}-{\bf Q}_{k,a,k',b}\right]}\!\right\}^{\!N}\nonumber\!\!\! ,
\end{align}
with the new normalization
\begin{align}
\mathcal{N}=\prod_{a=1}^{y_0}\prod_{k=0}^{k_f}\int_{-\infty}^{+\infty}d{w^{k,a}}e^{-\frac{{\bf w}{\bf Q}^{-1}{\bf w}}{2}}
\end{align}
and
\begin{align}
    {\bf w}=\left\{w^{k,a}\right\}_{k\in[\![1,\dots,k_f]\!],\,a\in[\![1,\dots,y_0]\!]}\, .
\end{align}
This completes the computation of the free energy in the chain setting -$y_k=1$ for all layer $k(>0)$-. To further simplify this expression, we will assume that the computation is \emph{annealed}. This means that configurations with different replica indices do not interact:
\begin{align}
         {\bf Q}_{k,a,{k'},b}=\delta_{a,b}\,  {\bf Q}_{k,{k'}}\, ,\\
         \hat{\bf Q}_{k,a,{k'},b}=\delta_{a,b}\,  \hat{\bf Q}_{k,{k'}}\, .
\end{align}
This yields directly

\begin{align}
    \tilde{\mathcal Z}=&\left\{\prod_{k=0}^{k_f}\int_{-\kappa}^{\kappa}d{w^{k}}\frac{e^{-\frac{{\bf w}{\bf Q}^{-1}{\bf w}}{2}}}{\mathcal{N}}\right\}^{y_0M}\\
    &\hspace{-0.8cm}\times\!\!\left\{\prod_{k=0}^{k_f}\sum_{{x}^{{k}}=\pm 1}e^{\underset{k,{k'}}{\sum}\!\hat{\bf Q}_{k,k'}\left[{x}^{{k}}{ x}^{{k'}}-{\bf Q}_{k,k'}\right]}\!\right\}^{y_0N}\nonumber\!\!\! ,
\end{align}
with
\begin{align}
\mathcal{N}=\prod_{k=0}^{k_f}\int_{-\infty}^{+\infty}d{w^{k}}e^{-\frac{{\bf w}{\bf Q}^{-1}{\bf w}}{2}}
\end{align}
and
\begin{align}
    {\bf w}=\left\{w^{k}\right\}_{k\in[\![1,k_f]\!]}\,.
\end{align}
\vspace{0.001cm}
This yields the \emph{annealed} free energy
\begin{align}
    \phi=&\!\!\underset{
    \tiny\begin{array}{c}
         \hat{\bf Q}_{k,k'}  \\
         {\bf Q}_{k,k'}\\
           ({\bf Q}_{k,k\pm1}\!=\!m)
    \end{array}}{\rm opt}\!\!\!\!\!\left\{\!N\log\!\left[\prod_{k=0}^{k_f}\sum_{{x}^{{k}}=\pm 1}e^{\underset{k,{k'}}{\sum}\!\hat{\bf Q}_{k,k'}\left({x}^{{k}}{ x}^{{k'}}-{\bf Q}_{k,k'}\right)}\!\right]\right.\nonumber\\
    &\hspace{1.2cm}+\!\!\left.M\log\!\left[\prod_{k=0}^{k_f}\int_{-\kappa}^{\kappa}d{w^{k}}\frac{e^{-\frac{{\bf w}{\bf Q}^{-1}{\bf w}}{2}}}{\mathcal{N}}\right]\!\right\} .
\end{align}

\section{The coarse-graining approach}
\label{app: coarse-graining}
Taking the limit $m\rightarrow 1$, we can assume that the overlap and field matrices ${\bf Q}$ and $\hat{\bf Q}$ can be coarse-grained. To do so, we will consider that a large number of the entries of ${\bf Q}^{-1}$ and $\hat{\bf Q}$ can be set to zero (without losing precision for the saddle point evaluation). We define a subgrid of configurations 
\begin{align}
    {\bf x}^*&=\left\{x^{*,k}\right\}_{k\in\left[\!\left[1,k_f^*\right]\!\right]}\,,\\
    {\bf w}^*&=\left\{w^{*,k}\right\}_{k\in\left[\!\left[1,k_f^*\right]\!\right]}\,,
\end{align}
for which the interactions remain generic. The remaining configurations will follow the \emph{no-memory Ansatz} introduced in \cite{barbier2025findingrightpathstatistical}. We chose this $Ansatz$ because it is the simplest form for ${\bf Q}^{-1}$ and $\hat{\bf Q}$ that can ensure the connectivity criteria ${\bf x}^k\cdot{\bf x}^{k\pm1}/N=m$. We recall that it corresponds to taking for a no-memory element $x^k$ (respectively $w^k$)
\begin{align}
    {\bf Q}^{-1}_{k,k'(\neq k)}&=\delta_{k',k-1}{\bf Q}^{-1}_{k,k-1}+\delta_{k',k+1}{\bf Q}^{-1}_{k,k+1}\, ,\\
    \hat{\bf Q}_{k,k' (\neq k)}&=\delta_{k',k-1}\hat{\bf Q}_{k,k-1}+\delta_{k',k+1}\hat{\bf Q}_{k,k+1}\, .
\end{align}
As we will see, this $Ansatz$ allows us to coarse-grain out all configurations except for ${\bf x}^*$ and ${\bf w}^*$. By doing so, we will be left with simulating chains with only $k_f^*$ elements. This will allow us to evaluate the saddle point for the free energy in the limit $m\rightarrow1$ without having to perform heavy numerical calculations. 

In the following, we will assume for simplicity that we have $N_0$ no-memory configurations $\{x^{j,k}\}_{j\in[\![1,N_0]\!]}$ (respectively, $\{w^{j,k}\}_{j\in[\![1,N_0]\!]}$) between two configurations $x^{*,k}$ and $x^{*,k+1}$ (respectively, $w^{*,k}$ and $w^{*,k+1}$).

\subsection{entropic contribution}
\label{app: coarse-graining entropic contribution}
We focus on the coarse-graining scheme for the entropic contribution. Under the no-memory $Ansatz$ we have
\begin{align}
    &\prod_{k=0}^{k_f}\sum_{{x}^{{k}}=\pm 1}e^{\underset{k,{k'}}{\sum}\!\hat{\bf Q}_{k,k'}\left[{x}^{{k}}{ x}^{{k'}}-{\bf Q}_{k,k'}\right]}=\\
    &\prod_{k=0}^{k_f^*}\sum_{{x}^{{*,k}}=\pm 1}\left(\prod_{j=1}^{N_0}\sum_{{x}^{{j,k}}=\pm 1}e^{h_k\left[x^{*,k}x^{1,k}-m\right]+h_k\left[x^{*,k+1}x^{N_0,k}-m\right]}\right.\nonumber\\
    &\hspace{1.3cm}\left.\times e^{\underset{j(\neq 1)}{\sum}h_k\left[x^{j,k}x^{j-1,k}-m\right]}\right)\ e^{\underset{k,{k'}}{\sum}\!\hat{\bf Q}^*_{k,k'}\left[{x}^{{*,k}}{ x}^{{*,k'}}-{\bf Q}^*_{k,k'}\right]}\,.\nonumber
\end{align}
The fields $\{h_k\}_{k\in[\![0,k_f^*-1]\!]}$ are the couplings that remain under the no-memory $Ansatz$. They ensure that we get an average overlap $m$ between two neighbors. Due to the properties of this $Ansatz$, these fields take a single value between two variables $x^{*,k}$ and $x^{*,k+1}$ -$h_k$ in this precise case-.
We can now note that the elements $\{x^{j,k}\}_{j\in[\![1,N_0]\!]}$ (for all $k$) take the form of a 1D Ising chain. They can be integrated using standard techniques such as transfer matrices \cite{book_Potters}, which yields
\begin{align}
    &\prod_{k=0}^{k_f}\sum_{{x}^{{k}}=\pm 1}e^{\underset{k,{k'}}{\sum}\!\hat{\bf Q}_{k,k'}\left[{x}^{{k}}{ x}^{{k'}}-{\bf Q}_{k,k'}\right]}=\\
    &\prod_{k=0}^{k_f^*}\sum_{{x}^{{*,k}}=\pm 1}\prod_{k=0}^{k_f^*-1}\left[{\bf t}^{N_0+1}(x^{*,k},x^{*,k+1}\vert h_k,m)\right]\nonumber\\
    &\hspace{1cm}\times e^{\underset{k,{k'}}{\sum}\!\hat{\bf Q}^*_{k,k'}\left[{x}^{{*,k}}{ x}^{{*,k'}}-{\bf Q}^*_{k,k'}\right]}\nonumber\,,
\end{align}
with
\begin{align}
    {\bf t}(x^{*,k},x^{*,k+1}\vert h_k,m)\!=\!\begin{pmatrix}
        e^{h_k-h_km}&e^{-h_k-h_km}\\
        e^{-h_k-h_km}&e^{h_k-h_km}
    \end{pmatrix}.
\end{align}

\subsection{energetic contribution}
\label{app: coarse-graining energetic contribution}
We now focus on the coarse-graining scheme for the energetic contribution. Under the no-memory $Ansatz$ we have
\begin{align}
    &\prod_{k=0}^{k_f}\int_{-\kappa}^{\kappa}d{w^{k}}{e^{-\frac{{\bf w}{\bf Q}^{-1}{\bf w}}{2}}}=\\
    &\prod_{k=0}^{k_f^*}\int_{-\kappa}^{\kappa}d{w^{*,k}}\left(\prod_{j=1}^{N_0}\int^\kappa_{-\kappa}dw^{j,k}e^{-\frac{{Q}^{1}_{k}{w^{*,k}}w^{1,k}}{2}}\right.\nonumber\\
    &\left.\hspace{0cm}\times e^{-\frac{{Q}^{1}_{k}{w^{*,k+1}}w^{N_0,k}+\underset{j(\neq1)}{\sum}{ Q}^{1}_{k}w^{j,k}w^{j-1,k}+\underset{j}{\sum}{Q}^{0}_{k}w^{j,k}w^{j,k}}{2}}\right)\nonumber\\
     &\times e^{-\frac{{\bf w}^*{\bf Q}^{\rm eff}{\bf w}^*}{2}}\nonumber\, .
\end{align}
Again, for the margins $\{w^{j,k}\}_{j\in[\![1,N_0]\!],k\in[\![0,k_f^*]\!]}$ following the no-memory $Ansatz$, the only remaining coupling terms are $\{Q^0_k,Q^1_k\}_{k\in[\![0,k_f^*-1]\!]}$. The rest of the margins $\{w^{*,k}\}_{k\in[\![0,k_f^*]\!]}$ have a generic coupling matrix ${\bf Q}^{\rm eff}$. Overall, these couplings ensure that we have
\begin{align}
   \left\langle w^{j,k}w^{j+1,k}\right\rangle&\!=\! \left\langle w^{k,*}w^{1,k}\right\rangle\!=\!\left\langle w^{k+1,*}w^{N_0,k}\right\rangle\!=\!m\, ,\\
   \left\langle (w^{j,k})^2\right\rangle&\!=\!\left\langle (w^{k,*})^2\right\rangle=1\,,\\
   \left\langle w^{k,*}w^{k',*}\right\rangle&\!=\!{\bf Q}^*_{k,k'}
\end{align}
with
\begin{align}
\label{eq: app measure energ}
    \langle O\rangle=\frac{\prod_{k=0}^{k_f}\int d{w^{k}}{\,O\,e^{-\frac{{\bf w}{\bf Q}^{-1}{\bf w}}{2}}}}{\prod_{k=0}^{k_f}\int d{w^{k}}{e^{-\frac{{\bf w}{\bf Q}^{-1}{\bf w}}{2}}}}\, .
\end{align}

To make this expression simpler, we can identify the no-memory sections as Brownian diffusion processes. In absence of $w^{*,k}$ and $w^{*,k+1}$, the elements $\{w^{j,k}\}_{j\in[\![1,N_0]\!]}$ generate a diffusion with an effective average overlap $m^{k}$ and norm ${\mathcal N}^{k}$:
\begin{align}
    &\prod_{j=1}^{N_0}\int^\kappa_{-\kappa}dw^{j,k}e^{-\frac{{Q}^{1}_{k}{w^{*,k}}w^{1,k}+{Q}^{1}_{k}{w^{*,k+1}}w^{N_0,k}}{2}}\\
    &\hspace{1cm}\times e^{-\frac{\underset{j(\neq1)}{\sum}{ Q}^{1}_{k}w^{j,k}w^{j-1,k}+\underset{j}{\sum}{Q}^{0}_{k}w^{j,k}w^{j,k}}{2}}\nonumber\\
    =&\prod_{j=1}^{N_0}\int^\kappa_{-\kappa}dw^{j,k}e^{-\frac{\left(w^{1,k}-\frac{m^{k}}{{\mathcal N}^{k}}w^{*,k}\right)^2}{2\left[{\mathcal N}^{k}-\frac{(m^{k})^2}{{\mathcal N}^{k}}\right]}+\frac{\left(\frac{m^{k}}{{\mathcal N}^{k}}w^{*,k}\right)^2}{2\left[{\mathcal N}^{k}-\frac{(m^{k})^2}{{\mathcal N}^{k}}\right]}}\nonumber\\
    &\times  e^{-\frac{\left(w^{*,k+1}-\frac{m^{k}}{{\mathcal N}^{k}}w^{N_0,k}\right)^2}{2\left[{\mathcal N}^{k}-\frac{(m^{k})^2}{{\mathcal N}^{k}}\right]}+\frac{\left(w^{*,k+1}\right)^2}{2\left[{\mathcal N}^{k}-\frac{(m^{k})^2}{{\mathcal N}^{k}}\right]}}\nonumber\\
    &\times e^{-\frac{\underset{j(\neq 1)}{\sum}\left(w^{j,k}-\frac{m^{k}}{{\mathcal N}^{k}}w^{j-1,k}\right)^2}{2\left[{\mathcal N}^{k}-\frac{(m^{k})^2}{{\mathcal N}^{k}}\right]}}\nonumber\\
    =&{\bf T}^{N_0+1}(w^{*,k+1},w^{*,k}\vert \kappa,m^{k},{\mathcal N}^{k}) e^{\frac{\left(w^{*,k+1}\right)^2+\left(\frac{m^{k}}{{\mathcal N}^{k}}w^{*,k}\right)^2}{2\left[{\mathcal N}^{k}-\frac{(m^{k})^2}{{\mathcal N}^{k}}\right]}}\nonumber
\end{align}
where we have the transfer matrix
\begin{align}
    {\bf T}(w,w'\vert \kappa,m,{\mathcal N})=\Theta(\kappa-\vert w\vert) e^{\frac{-\left(w-\frac{m}{\mathcal{N}}w'\right)^2}{2\left(\mathcal{N}-\frac{m^2}{\mathcal{N}}\right)}}\,.
\end{align}
This means that our previous computation can be rewritten as
\begin{align}
     &\prod_{k=0}^{k_f}\int_{-\kappa}^{\kappa}d{w^{k}}{e^{-\frac{{\bf w}{\bf Q}^{-1}{\bf w}}{2}}}=\\
     &\prod_{k=0}^{k_f^*}\int_{-\kappa}^{\kappa}d{w^{*,k}}\prod_{k=0}^{k_f^*-1}\left[{\bf T}^{N_0+1}(w^{*,k+1},w^{*,k}\vert \kappa,m^{k},{\mathcal N}^{k})\vphantom{e^{\frac{\left(w^{*,k+1}\right)^2+\left(\frac{m^{k}}{{\mathcal N}^{k}}w^{*,k}\right)^2}{2\left[{\mathcal N}^{k}-\frac{(m^{k})^2}{{\mathcal N}^{k}}\right]}}}\right.\nonumber\\
     &\hspace{1.3cm}\times \left.e^{\frac{\left(w^{*,k+1}\right)^2+\left(\frac{m^{k}}{{\mathcal N}^{k}}w^{*,k}\right)^2}{2\left[{\mathcal N}^{k}-\frac{(m^{k})^2}{{\mathcal N}^{k}}\right]}}\right]e^{-\frac{{\bf w}^*{\bf Q}^{\rm eff}{\bf w}^*}{2}}\nonumber
\end{align}

At this stage, we have to evaluate the different elements of ${\bf Q}^{\rm eff}$ and the no-memory couplings. For this, we have to estimate the previous expression in the limit $\kappa =\infty$ to retrieve the measure from Eq.~(\ref{eq: app measure energ}). In this limit, we have for the transfer matrix and its powers:
\begin{align}
        {\bf T}^n(w,w'\vert \kappa=+\infty,m,{\mathcal N})\propto e^{\frac{-\left(w-\frac{m_n}{\mathcal{N}}w'\right)^2}{2\left(\mathcal{N}-\frac{{m_n}^2}{\mathcal{N}}\right)}}
\end{align}
with $m_n=m^n/\mathcal{N}^{n-1}$ -and more generally $m^k_n=({m^k})^n/(\mathcal{N}^k)^{n-1}$-. This further simplifies the previous expression as 
\begin{align}
      &\prod_{k=0}^{k_f}\int d{w^{k}}{e^{-\frac{{\bf w}{\bf Q}^{-1}{\bf w}}{2}}}\propto\\
      &\prod_{k=0}^{k_f^*}\int d{w^{*,k}}\prod_{k=0}^{k_f^*-1}\left(e^{\frac{-\overset{k+1}{\underset{j,j'=k}{\sum}}w^{*,j}Q^k_{j,j'}w^{*,j'}}{2}}\right)e^{-\frac{{\bf w}^*{\bf Q}^{\rm eff}{\bf w}^*}{2}}\nonumber
\end{align}
with
\begin{align}
   {Q}_{k+1,k+1}^{k} &={Q}_{k,k}^{k}\\&=\frac{\left(\frac{m_{N_0+1}^k}{{\mathcal{N}^k}}\right)^2-\left(\frac{m^k}{{\mathcal{N}^k}}\right)^2}{\mathcal{N}^k\!\left[1\!-\!\left(\frac{m_{N_0+1}^k}{{\mathcal{N}^k}}\right)^2\right]\!\!\left[1\!-\!\left(\frac{m^k}{{\mathcal{N}^k}}\right)^2\right]}\, ,\nonumber\\
    {Q}_{k,k+1}^{k}&={Q}_{k+1,k}^{k}=-\frac{m_{N_0+1}^k}{\mathcal{N}^k\left[1\!-\!\left(\frac{m_{N_0+1}^k}{{\mathcal{N}^k}}\right)^2\right]}\, .
\end{align}
To obtain $\left\langle w^{k,*}w^{k',*}\right\rangle\!=\!{\bf Q}^*_{k,k'}$, we simply need to set
\begin{align}
    \left[{\bf Q}^{\rm eff}\right]_{k,k'}=&\left[{\bf Q}^{*-1}\right]_{k,k'}\\
    &-\delta_{k+1,k'}(1-\delta_{k,k_f^*}){Q}_{k,k+1}^{k}\nonumber\\
    &-\delta_{k-1,k'}(1-\delta_{k,1}){Q}_{k,k-1}^{k-1}\nonumber\\
    &-\delta_{k,k'}(1-\delta_{k,k_f^*}){Q}_{k,k}^{k}\nonumber\\
    &-\delta_{k,k'}(1-\delta_{k,1}){Q}_{k,k}^{k-1}\, .\nonumber
\end{align}
It remains to determine the terms $\{m^k,\mathcal{N}^k\}_{k\in[\![0,k_f^*-1]\!]}$. For this, we simply need to keep the first element $w^{1,k_0}$ -in the no-memory chain between $w^{*,k_0}$ and $w^{*,k_0+1}$- and evaluate its average norm and overlaps. More precisely, we recall that we have
\begin{align}
      &\prod_{k=0}^{k_f}\int d{w^{k}}{e^{-\frac{{\bf w}{\bf Q}^{-1}{\bf w}}{2}}}\\
      =&\prod_{k=0}^{k_f^*}\int d{w^{*,k}}\prod_{k=0}^{k_f^*-1}\left[{\bf T}^{N_0+1}(w^{*,k+1},w^{*,k}\vert \kappa,m^{k},{\mathcal N}^{k})\vphantom{e^{\frac{\left(w^{*,k+1}\right)^2+\left(\frac{m^{k}}{{\mathcal N}^{k}}w^{*,k}\right)^2}{2\left[{\mathcal N}^{k}-\frac{(m^{k})^2}{{\mathcal N}^{k}}\right]}}}\right.\nonumber\\
     &\hspace{1.3cm}\times \left.e^{\frac{\left(w^{*,k+1}\right)^2+\left(\frac{m^{k}}{{\mathcal N}^{k}}w^{*,k}\right)^2}{2\left[{\mathcal N}^{k}-\frac{(m^{k})^2}{{\mathcal N}^{k}}\right]}}\right]e^{-\frac{{\bf w}^*{\bf Q}^{\rm eff}{\bf w}^*}{2}}\nonumber\, .
\end{align}
After integrating all variables except for $w^{*,k_0}$, $w^{*,k_0+1}$ and $w^{1,k_0}$ we obtain directly
\begin{align}
    &\prod_{k=0}^{k_f}\int d{w^{k}}{e^{-\frac{{\bf w}{\bf Q}^{-1}{\bf w}}{2}}}\\
    \propto&\int d{w^{*,k_0}}d{w^{*,k_0+1}} dw^{1,k_0} e^{-\frac{\tilde{\bf w}\tilde{\bf Q}\tilde{\bf w}}{2}}\nonumber
\end{align}
with 
\begin{align}
    \tilde{\bf w}=\left[w^{*,k_0},w^{1,k_0},w^{*,k_0+1}\right]\,,
\end{align}
\begin{align}
    \tilde{\bf Q}&=\\
   &\hspace{-0.3cm}{\tiny \begin{pmatrix}
        A &\frac{-m^{k_0}}{\mathcal{N}^{k_0}\left[{\mathcal N}^{k_0}-\frac{\left(m^{k_0}\right)^2}{{\mathcal N}^{k_0}}\right]}  &B\\
         \frac{-m^{k_0}}{\mathcal{N}^{k_0}\left[{\mathcal N}^{k_0}-\frac{\left(m^{k_0}\right)^2}{{\mathcal N}^{k_0}}\right]}& C&\frac{-m_{N_0}^{k_0}}{\mathcal{N}^{k_0}\left[{\mathcal N}^{k_0}-\frac{\left(m_{N_0}^{k_0}\right)^2}{{\mathcal N}^{k_0}}\right]}\\
         B& \frac{-m_{N_0}^{k_0}}{\mathcal{N}^k\left[{\mathcal N}^{k_0}-\frac{\left(m_{N_0}^{k_0}\right)^2}{{\mathcal N}^{k_0}}\right]} &D\\
    \end{pmatrix}}\nonumber
\end{align}
and 
\begin{align}
    C=\frac{1}{\left[{\mathcal N}^{k_0}-\frac{\left(m^{k_0}\right)^2}{{\mathcal N}^{k_0}}\right]}+\frac{\left(\frac{m_{N_0}^{k}}{{\mathcal N}^{k_0}}\right)^2}{\left[{\mathcal N}^{k_0}-\frac{\left(m_{N_0}^{k_0}\right)^2}{{\mathcal N}^{k_0}}\right]}\,.
\end{align}
The variables $A$, $B$ and $D$ will be determined in the following. Finally, we can set $m^k$ and $\mathcal{N}^k$ by verifying the correct correlations, i.e.
\begin{align}
    {\tilde{\bf Q}}^{-1}=\begin{pmatrix}
        1&m&{\bf Q}^*_{k_0,k_0+1}\\
        m&1&E\\
        {\bf Q}^*_{k_0+1,k_0}&E&1
    \end{pmatrix}\, .
\end{align}
This corresponds to a set of 6 equations to be solved with $A$,$B$,$D$,$E$, $m^{k_0}$ and $\mathcal{N}^{k_0}$ as variables. After all these steps, we have fixed ${\bf Q}^{\rm eff}$ and the no-memory parameters $\{m^k,\mathcal{N}^k\}_{k\in[\![1,k_f^*-1]\!]}$.

\subsection{The coarse-grained free energy and the saddle point equations}
\label{app: coarse-graining saddle-point}

With the previous section, we have fixed all the parameters for the coarse-graining of the free energy. This means that we now have
\begin{align}
    \phi=&N\log\left\{\prod_{k=0}^{k_f^*}\sum_{{x}^{{*,k}}=\pm 1} e^{\underset{k,{k'}}{\sum}\!\hat{\bf Q}^*_{k,k'}\left[{x}^{{*,k}}{ x}^{{*,k'}}-{\bf Q}^*_{k,k'}\right]}  \right.\\
    &\left. \hspace{2cm}\times \prod_{k=0}^{k_f^*-1}e^{\varphi_S(x^{*,k+1},x^{*,k}\vert h_{k},m)}\right\}\nonumber\\
    &+M\log\left\{\prod_{k=0}^{k_f^*}\int_{-\kappa}^\kappa dw^{*,k}e^{-\frac{{\bf w}^*{\bf Q}^{\rm eff}{\bf w}^*}{2}}\right.\nonumber\\
    &\left.\hspace{1.5cm}\times\prod_{k=0}^{k_f^*-1}\left[e^{\varphi_E(w^{*,k+1},w^{*,k}\vert \kappa,m^{k},{\mathcal N}^{k})}\right]\right\}\nonumber\\
     &-M\log\left\{\prod_{k=0}^{k_f^*}\int dw^{*,k}e^{-\frac{{\bf w}^*{\bf Q}^{\rm eff}{\bf w}^*}{2}}\right.\nonumber\\
    &\left.\hspace{1.2cm}\times\prod_{k=0}^{k_f^*-1}\left[e^{\varphi_E(w^{*,k+1},w^{*,k}\vert \kappa=+\infty,m^{k},{\mathcal N}^{k})}\right]\right\}\nonumber
\end{align}
with
\begin{align}
&\varphi_S(x^{*,k+1},x^{*,k}\vert h_{k},m)=\\
    &\hspace{1cm}\log\left[{\bf t}^{N_0+1}(x^{*,k},x^{*,k+1}\vert h_k,m)\right]\nonumber
\end{align}
and 
\begin{align}
&\varphi_E(w^{*,k+1},w^{*,k}\vert \kappa,m^{k},{\mathcal N}^{k})={\frac{\left(\!w^{*,k+1}\right)^{\!2}\!+\!\left(\!\frac{m^{k}}{{\mathcal N}^{k}}w^{*,k}\!\right)^{\!2}}{2\left[{\mathcal N}^{k}-\frac{(m^{k})^2}{{\mathcal N}^{k}}\right]}}\nonumber\\
    &\hspace{1.3cm}+\log\!\left\{\!{\bf T}^{\small {N_0+1}}(w^{*,k+1},w^{*,k}\vert \kappa,m^{k},{\mathcal N}^{k}) \!\right\}.
\end{align}
The saddle-point equations we now have to verify are
\begin{align}
    \label{eq: s-p 1}
    &\frac{\partial \phi}{\partial \hat{\bf Q}^*_{k,k'}}=0
    \iff{\bf Q}^*_{k,k'}=\left\langle x^{*,k}x^{*,k'}\right\rangle_S,\\
    \label{eq: s-p 2}
    &\frac{\partial \phi}{\partial h_k}\!=\!0\!\iff\!0\!=\!\left\langle\frac{\partial\varphi_S(x^{*,k+1},x^{*,k}\vert h_{k},m)}{\partial h_k}\right\rangle_{\!S},
    \end{align}
and (with $k'>k$)
\begin{align}
    \label{eq: s-p 3}
     &\frac{\partial \phi}{\partial {\bf Q}^*_{k,k'}}=0
    \iff\hat{\bf Q}^*_{k,k'}=\\
    &+\delta_{k',k+1}\alpha\left\langle\frac{\delta\varphi_E(w^{*,k+1},w^{*,k}\vert \kappa,m^{k},{\mathcal N}^{k})}{\partial {\bf Q}^*_{k,k'} }\right\rangle_{E,\kappa}\nonumber\\
    &-\delta_{k',k+1}\alpha\left\langle\frac{\delta\varphi_E(w^{*,k+1},w^{*,k}\vert \kappa,m^{k},{\mathcal N}^{k})}{\partial {\bf Q}^*_{k,k'} }\right\rangle_{E,\kappa=+\infty}\nonumber\\
    &-\frac{\alpha}{2}\sum_{j,j'}\frac{\partial {\bf Q}^{\rm eff}_{j,j'}}{\partial {\bf Q}^*_{k,k'}}\!\left[\!\left\langle \! w^{*,j}w^{*,j'}\right\rangle_{E,\kappa}\!\!-\!\left\langle \!w^{*,j}w^{*,j'}\right\rangle_{E,\kappa=+\infty}\right]\nonumber\!.
\end{align}
We used here the compact notations:
\begin{align}
\label{eq: measure 1}
    &\langle O\rangle_S=\\
    &\frac{\overset{k_f^*}{\underset{k=0}{\prod}}\underset{{{x}^{{*,k}}=\pm 1}}{\sum}\! O \,e^{\underset{k,{k'}}{\sum}\!\hat{\bf Q}^*_{k,k'}{x}^{{*,k}}{ x}^{{*,k'}}+\underset{k}{\sum} \varphi_S(x^{*,k+1},x^{*,k}\vert h_{k},m)}}{\overset{k_f^*}{\underset{k=0}{\prod}}\underset{{{x}^{{*,k}}=\pm 1}}{\sum}  e^{\underset{k,{k'}}{\sum}\!\hat{\bf Q}^*_{k,k'}{x}^{{*,k}}{ x}^{{*,k'}}+\underset{k}{\sum} \varphi_S(x^{*,k+1},x^{*,k}\vert h_{k},m)}}\nonumber
\end{align}
and
\begin{align}
\label{eq: measure 2}
    &\langle O\rangle_{E,\kappa}=\\
    &\frac{\overset{k_f^*}{\underset{k=0}{\prod}}\int_{-\kappa}^\kappa dw^{*,k} O e^{-\frac{{\bf w}^*{\bf Q}^{\rm eff}{\bf w}^*}{2}+\underset{k}{\sum} \varphi(w^{*,k+1},w^{*,k}\vert \kappa,m^{k},{\mathcal N}^{k})}}{\overset{k_f^*}{\underset{k=0}{\prod}}\int_{-\kappa}^\kappa dw^{*,k}  e^{-\frac{{\bf w}^*{\bf Q}^{\rm eff}{\bf w}^*}{2}+\underset{k}{\sum} \varphi(w^{*,k+1},w^{*,k}\vert \kappa,m^{k},{\mathcal N}^{k})}}.\nonumber
\end{align}
To solve these equations, we first set a guess for ${\bf Q}^*$, $\hat{\bf Q}^*$ and $\{h_k\}_{k\in[\![0,k_f^*-1]\!]}$. We then evaluate each observable involved in the right-hand side of Eqs.~(\ref{eq: s-p 1},\ref{eq: s-p 2},\ref{eq: s-p 3}) by Monte Carlo sampling the measures from Eqs.~(\ref{eq: measure 1},\ref{eq: measure 2}). This returns a new estimate for ${\bf Q}^*$ and $\hat{\bf Q}^*$; the fields $\{h_k\}_{k\in[\![0,k_f^*-1]\!]}$ are updated until we find a solution for Eq.~(\ref{eq: s-p 2}). When the input guess and the returned estimates match, the saddle point is found.

\bibliography{apssamp}
\end{document}